\documentclass[twocolumn, floatfix, prl, aps, showpacs, longbibliography, 10pt]{revtex4-2}
\usepackage{graphicx, amssymb, color, amsmath}
\usepackage{nicefrac}
\usepackage{multirow, array, booktabs}
\usepackage{longtable}
\usepackage{float}
\usepackage{romannum}
\usepackage{mathtools}
\usepackage{bbm}
\usepackage{mathrsfs}
\usepackage{IEEEtrantools}
\usepackage{relsize}
\usepackage{mathrsfs}
\usepackage{mathtools}
\usepackage{xparse}
\usepackage[titletoc, title]{appendix}
\usepackage[colorlinks, bookmarks=true, citecolor=blue, linkcolor=red, urlcolor=blue]{hyperref}
\usepackage{orcidlink}
\AtBeginDocument{\pagenumbering{arabic}}

\graphicspath{{./figures/}}
\begin{document}

\title{Relations between density-density correlators of states in the maximal spin multiplet}

\author{Ritajit Kundu\orcidlink{0000-0003-4917-4020}}
\email{ritajitk@imsc.res.in}
\affiliation{Institute of Mathematical Sciences, CIT Campus, Chennai, 600113, India}

\author{Ajit C. Balram\orcidlink{0000-0002-8087-6015}}
\email{cb.ajit@gmail.com}
\affiliation{Institute of Mathematical Sciences, CIT Campus, Chennai 600113, India}
\affiliation{Homi Bhabha National Institute, Training School Complex, Anushakti Nagar, Mumbai 400094, India} 

\begin{abstract}
    We present identities relating the pair-correlation functions and static structure factors of states in the maximal spin multiplet. This allows us to compute these density-density correlation functions of all members of the multiplet using just these correlation functions of the highest-weight state. We apply these relations to obtain energies for many fractional quantum Hall (FQH) states. In particular, we analytically compute the energies of the Halperin-$(1,1,1)$ state as a function of density imbalance and layer separation, and numerically evaluate these energies for many other FQH states. 
\end{abstract}

\maketitle 

\textbf{\textit{Introduction}.} 
Correlation functions provide a bridge between the microscopic structure and measurable properties in many-body systems \cite{Kubo57}. The pair-correlation function, $g(r)$, encodes the spatial correlations in a translation and rotational invariant state by specifying the probability of finding two particles a distance $r$ apart from each other in it~\cite{Hansen13}. Its Fourier transform, the static structure factor $S(q)$ at momentum $q$, is directly accessible in scattering experiments~\cite{Friedrich13, Franklin53, Ashcroft66, Greenfield71, Yarnell73, Vahvaselk78, Svensson80, Furrer09} and governs key physical properties, including interaction energies and linear response~\cite{Giuliani08, Mahan13}. Together, these density-density correlators, $g(r)$ and $S(q)$, serve as primary diagnostics of correlation effects in many systems~\cite{Giuliani08, Pines18}.

Yet computing $g(r)$ or $S(q)$ in strongly correlated systems remains a significant challenge. Exact diagonalization is controlled but restricted to small system sizes, making thermodynamic extrapolation from it unreliable \cite{Yoshioka83, Lin90, Pereira07, Fehske07, Zhang10}. Monte Carlo methods~\cite{Binder10, Fehske07} access larger systems but can become computationally expensive when exploring a broad regime of parameter space. The problem is compounded in systems with spin or pseudospin (i.e., layer, orbital, or valley) degrees of freedom: a state with total spin $\mathbb{S}$ forms a $(2\mathbb{S}{+}1)$-fold multiplet whose members, related by spin-raising and spin-lowering operators, all have distinct correlation functions. As a result, properties across the multiplet have traditionally been evaluated separately for each member, further increasing the computational cost by a factor of $(2\mathbb{S}{+}1)$.

In this Letter, we show that the $g(r)$ and $S(q)$ of states within the maximal spin multiplet are not independent but satisfy exact relations that are determined solely by the $SU(2)$-spin rotation symmetry connecting the members of the multiplet. These relations enable the evaluation of energies and other properties of all states in the multiplet to be obtained from that of a single reference state, which eliminates the need for separate computations for each state in the multiplet. We demonstrate the power of this approach by first analytically computing the exact energy of the interlayer exciton condensate state that occurs in the fractional quantum Hall (FQH) regime. We then numerically evaluate energies of a broad class of FQH states across a wide range of parameters. Owing to its generality, our method provides a versatile framework for studying correlated quantum systems.

\textbf{\textit{Relations between density-density correlation functions in the maximal spin multiplet}.} The pair-correlation function for a translation and rotational invariant normalized state $|\Psi\rangle$ with $N$ particles with density $\rho_{0}$ is defined as
\begin{eqnarray}
\label{eq: define_pair_correlation}
g^{\sigma, \sigma'}\left(r=|\vec{r}|\right)&=&\frac{N(N-1)}{\rho_0^2}\int d^{D}
\vec{r}_{3}{\cdots}d^{D}\vec{r}_{N} {\times}  \\
&&|\Psi(\{\vec{r},\sigma\}, \{\vec{0},\sigma'\}, \{\vec{r}_{3},\sigma_{3}\},{\cdots},\{\vec{r}_{N},\sigma_{N}\} )|^{2},\nonumber 
\end{eqnarray}
where $D$ is the dimension. In words, the spin-resolved pair-correlation function is obtained by integrating out all other particles from the $N{-}$particle density matrix, $|\Psi|^{2}$, except for the two particles, chosen to lie at position $\vec{r}$ and the origin $\vec{0}$ with specified spins $\sigma$ and $\sigma'$. 

Without loss of generality, we take the reference state to be the highest-weight state with $\mathbb{S}_{z}{=}\mathbb{S}$. We consider a fully polarized translation and rotational invariant state of $N$ spin-$1/2$ particles with $\mathbb{S}{=}\mathbb{S}_{z}{=}N/2$, whose pair-correlation function, $g(r)$, is given. Using it, we can obtain the pair-correlation function of a state within the same multiplet, i.e., a state with $\mathbb{S}{=}N/2,~\mathbb{S}_{z}{<}N/2$, as
\begin{eqnarray}
 \label{eq: pair_correlation_Sz_multiplet}
    g^{\uparrow, \uparrow}\left(r\right)&=&\frac{N_{\uparrow}(N_{\uparrow}-1)}{N(N-1)}g(r),~
    g^{\downarrow, \downarrow}\left(r\right)=\frac{N_{\downarrow}(N_{\downarrow}-1)}{N(N-1)}g(r),\nonumber\\
    g^{\uparrow, \downarrow}\left(r\right)&=&\frac{N_{\uparrow}N_{\downarrow}}{N(N-1)}g(r)=g^{\downarrow, \uparrow}\left(r\right),
\end{eqnarray}
where $N_{\uparrow}$ and $N_{\downarrow}$ are the number of particles with $\uparrow$ and $\downarrow$ spins, respectively, and we assume without any loss of generality that $N_{\uparrow}{\geq}N_{\downarrow}$. In other words, the $g(r)$ of the fully polarized state splits into its various components, i.e., $g\left(r\right){=}g^{\uparrow, \uparrow}\left(r\right){+}g^{\uparrow, \downarrow}\left(r\right){+}g^{\downarrow, \uparrow}\left(r\right){+}g^{\downarrow, \downarrow}\left(r\right)$, each of which is the $g(r)$ times the appropriate number of pairs of particles of that component. The $g^{\uparrow, \uparrow}\left(r\right)$ is weighted by the number of ${\uparrow}{-}{\uparrow}$ pairs over the total pairs, i.e, $\binom{N_{\uparrow}}{2}/\binom{N}{2}$; $g^{\downarrow, \downarrow}\left(r\right)$ is weighted by the number of ${\downarrow}{-}{\downarrow}$ pairs over the total pairs, i.e, $\binom{N_{\downarrow}}{2}/\binom{N}{2}$, and $g^{\uparrow, \downarrow}\left(r\right)$ and $g^{\downarrow, \uparrow}\left(r\right)$ are weighted by half the number of ${\uparrow}{-}{\downarrow}$ pairs over the total pairs, i.e, $(1/2)N_{\uparrow}N_{\downarrow}/\binom{N}{2}$. The corresponding relations in the thermodynamic limit of $N_{\uparrow},N_{\downarrow}{\gg}1$, are
\begin{eqnarray}
 \label{eq: pair_correlation_Sz_multiplet_thermodynamic}
    g^{\uparrow, \uparrow}\left(r\right)&=&\frac{N_{\uparrow}^2}{N^2}g(r),~
    g^{\downarrow, \downarrow}\left(r\right)=\frac{N_{\downarrow}^2}{N^2}g(r),\nonumber\\
    g^{\uparrow, \downarrow}\left(r\right)&=&\frac{N_{\uparrow}N_{\downarrow}}{N^2}g(r)=g^{\downarrow, \uparrow}\left(r\right).
\end{eqnarray}
The translation and rotational invariance of the highest-weight state gets inherited by all members of the multiplet. Moreover, the relations of Eq.~\eqref{eq: pair_correlation_Sz_multiplet_thermodynamic} show that the vanishing properties (if there are any), i.e., how $g(r){\to}0$ as $r{\to}0$, for all members of the multiplet, are identical, with only the prefactor differing for different members of the multiplet. 

An analogous relation for the static structure factor is
\begin{eqnarray}
 \label{eq: structure_factor_Sz_multiplet_thermodynamic}
    S^{\sigma, \sigma'}\left(q\right)&=&\frac{\delta_{\sigma, \sigma'}N_{\sigma}}{N}+\rho_{0}\int d^{D}\vec{r} e^{-i \vec{q}\cdot \vec{r}}\left[g^{\sigma, \sigma'}(r)-\frac{N_{\sigma}N_{\sigma'}}{N^2}\right], \nonumber \\
    S^{\uparrow, \uparrow}\left(q\right)&{=}&\frac{N_{\uparrow}}{N}{+}\frac{N_{\uparrow}^{2}}{N^{2}}\left[S(q)-1\right], \nonumber \\
    S^{\downarrow, \downarrow}\left(q\right)&{=}&\frac{N_{\downarrow}}{N}{+}\frac{N_{\downarrow}^2}{N^2}\left[S(q)-1\right],\\
    S^{\uparrow, \downarrow}\left(q\right)&{=}&\frac{N_{\uparrow}N_{\downarrow}}{N^2}\left[S(q)-1\right]=S^{\downarrow, \uparrow}\left(q\right).\nonumber
\end{eqnarray}

The energies of the various states within the same maximal spin multiplet can now be obtained using Eq.~\eqref{eq: pair_correlation_Sz_multiplet_thermodynamic} or~\eqref{eq: structure_factor_Sz_multiplet_thermodynamic}, with input of only the $g(r)$ or $S(q)$ of the highest-weight state in the multiplet. The identity given in Eq.~\eqref{eq: pair_correlation_Sz_multiplet_thermodynamic}, and the closely related Eq.~\eqref{eq: structure_factor_Sz_multiplet_thermodynamic} are the central results of our work. We emphasize that these relations only apply to states within the maximal spin multiplet, i.e., all members of which have the same $\mathbb{S}{=}N/2$, but have different $\mathbb{S}_{z}$, and do not relate states across different multiplets, i.e., those with different $\mathbb{S}$. Next, we will use Eq.~\eqref{eq: pair_correlation_Sz_multiplet_thermodynamic} to compute energies of many FQH states as a function of certain parameters. 

\textbf{\textit{Energies from density-density correlations}.}
The pair-correlation function can be used to compute the energy of the state in the thermodynamic limit as~\cite{Giuliani08}
\begin{eqnarray}
\label{eq: energy_background_subtracted_per_particle_pair_correlation_thermodynamic}
    E_{0} &=& \frac{\rho_0}{2} \sum_{\sigma, \sigma'} \int d^{D}\vec{r}~v^{\sigma, \sigma'}(r,d)\left[g^{\sigma, \sigma'}(r)-\frac{N_{\sigma}N_{\sigma'}}{N^2}\right], \\
    &=& \frac{\nu}{2} \sum_{\sigma, \sigma'} \int_{0}^{\infty} dr~r v^{\sigma, \sigma'}(r,d)\left[g^{\sigma, \sigma'}(r)-\frac{N_{\sigma}N_{\sigma'}}{N^2}\right].\nonumber 
\end{eqnarray}
In the last line, we have restricted to two dimensions, considering an FQH state at filling $\nu{=}2\pi\ell^{2}\rho_{0}$, where $\ell{=}\sqrt{\hbar c/(eB)}$ is the magnetic length at magnetic field $B$. From here on, we will consider bilayer FQH states, where the spin index stands for the layer degree of freedom. In this setting, the interaction $v^{\uparrow,\uparrow}(r){=}1/r{=}v^{\downarrow,\downarrow}(r)$, and $v^{\uparrow,\downarrow}(r,d){=}1/\sqrt{r^2{+}d^{2}}{=}v^{\downarrow,\uparrow}(r,d)$, where $d$ is the layer separation in units of $\ell$ (for ease of notation, we will set $\ell{=}1$ in many instances). This interaction breaks the $SU(2)$-spin rotation symmetry, but the states we consider all have a definitive spin quantum number $\mathbb{S}{=}N/2$. Thus, our calculations are variational in nature, and the states can only provide a good microscopic representation of the actual ground state, such as the one obtained from exact diagonalization, in the $d{\ll}\ell$ limit. All the energies are quoted in Coulomb units of $e^{2}/(\varepsilon\ell)$, where $\varepsilon$ is the dielectric constant of the background host material. 

Analogous expression for the energy in terms of the static structure factor is
\begin{eqnarray}
\label{eq: energy_background_subtracted_per_particle_structure_factor_thermodynamic}
    E_{0} &{=}& \frac{1}{2} \sum_{\sigma, \sigma'} \int \frac{d^{D}\vec{q}}{(2\pi)^{D}}~\tilde{v}^{\sigma, \sigma'}(q,d)\left[S^{\sigma, \sigma'}(q){-}\delta_{\sigma, \sigma'}\frac{N_{\sigma}}{N}\right], \\
    &{=}& \frac{1}{4\pi} \sum_{\sigma, \sigma'} \int_{0}^{\infty} dq~q \tilde{v}^{\sigma, \sigma'}(q,d)\left[S^{\sigma, \sigma'}(q){-}\delta_{\sigma, \sigma'}\frac{N_{\sigma}}{N}\right]\nonumber ,
\end{eqnarray}
where the last line is again for $D{=}2$ dimensions, and $\tilde{v}^{\sigma, \sigma'}(q,d)$ is the Fourier transform of the interaction $v^{\sigma, \sigma'}(r,d)$. For $v(r,d){=}1/\sqrt{r^2{+}d^{2}}~e^{2}/(\varepsilon\ell)$, $\tilde{v}(q,d){=}(2\pi/q)e^{-q d}~e^{2}/(\varepsilon\ell)$. We will consider states with polarization $\gamma{=}(N_{\uparrow}{-}N_{\downarrow})/(N_{\uparrow}{+}N_{\downarrow})$, which also equals the relative filling imbalance $\delta\nu/\nu{\equiv}(\nu_{\uparrow}{-}\nu_{\downarrow})/(\nu_{\uparrow}{+}\nu_{\downarrow})$, where $N_{\uparrow}{=}N/2{+}\mathbb{S}_{z}$ and $N_{\downarrow}{=}N/2{-}\mathbb{S}_{z}$, with $\mathbb{S}_{z}{=}N\gamma/2$, and compute their energies next. Note that $0{\leq}\gamma{\leq}1$.

\textbf{\textit{Results: applications to fractional quantum Hall fluids}.} 
Since all states in the $\mathbb{S}{=}N/2$ multiplet have the maximal total spin, the spin part of the wavefunction is fully symmetric~\cite{Dora25}, and will be omitted in what follows for ease of notation. The spatial part of the wavefunction is obtained from the spatial part of the highest-weight state by relabeling $N_{\downarrow}$ of the coordinates in it as $\downarrow$ particles. This spatial part of the wavefunction also shows that the vanishing properties of all members of a multiplet are identical up to prefactors. Furthermore, this prefactor gets fixed by noting that the pair-correlation functions defined in Eq.~\eqref{eq: define_pair_correlation} are normalized such that as $r{\to}\infty$, $g^{\sigma, \sigma'}\left(r\right){\to}N_{\sigma}N_{\sigma'}/N^2$~\cite{Dora25}, which is the probability of finding two far-separated particles with spin-$\sigma$ and spin-$\sigma'$, which is just proportional to the product of their densities since as $r{\to}\infty$ the positions of the particles become independent of each other. This provides an alternative way to obtain Eq.~\eqref{eq: pair_correlation_Sz_multiplet_thermodynamic}. This discussion also demonstrates why simple linear relations like that given in Eq.~\eqref{eq: pair_correlation_Sz_multiplet} do not exist for states in a non-fully polarized multiplet. This is because in non-fully polarized states, the spin part of the wavefunction is intertwined with the real-space part and cannot be decoupled, as in the case of fully polarized states~\cite{Dora25}.

\begin{figure*}[tbh!]
\includegraphics{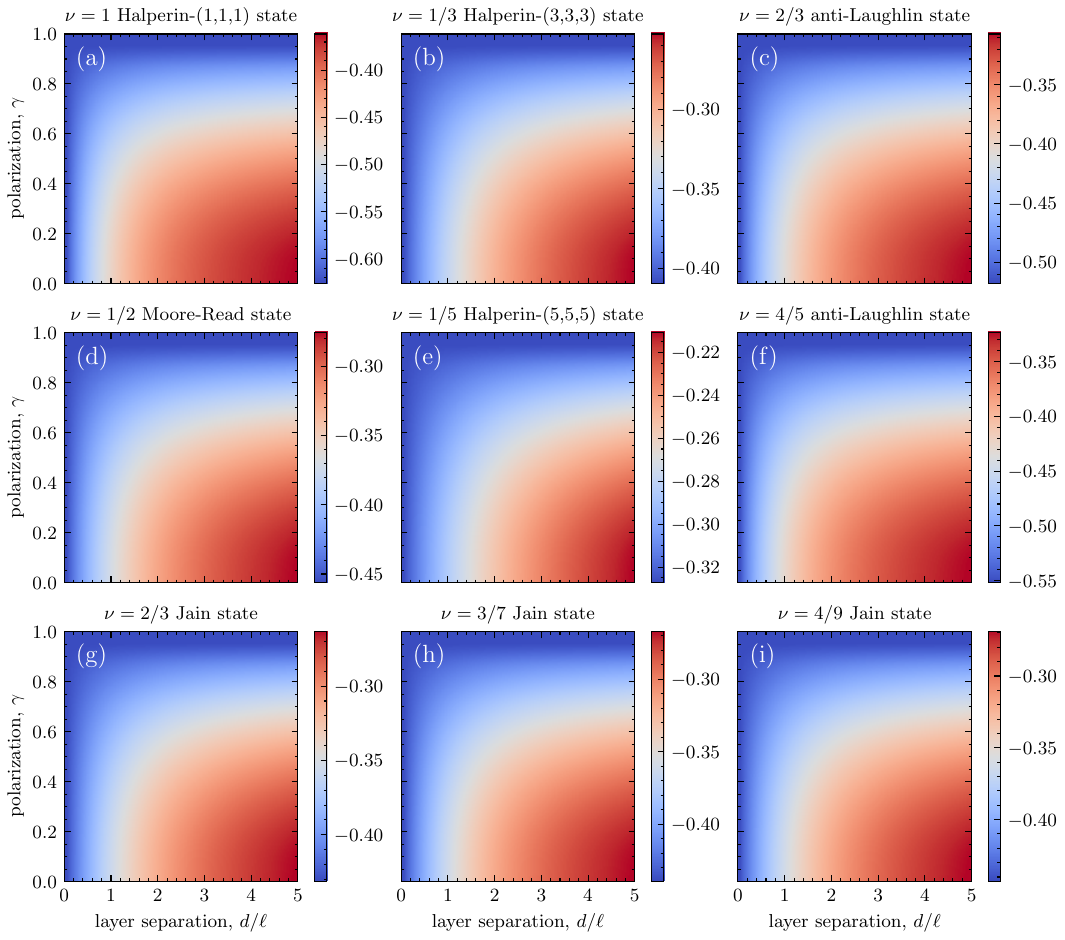}
          \caption{Coulomb energies in units of $e^{2}/(\varepsilon\ell)$ in the thermodynamic limit of many fully polarized, i.e., with maximal total spin, fractional quantum Hall states as a function of layer separation, $d/\ell$, and polarization, $\gamma$, which labels the $z{-}$component of the spin. Here, $\ell$ is the magnetic length and $\varepsilon$ is the dielectric constant of the background host material.}
          \label{fig: Coulomb_energy_FQH_states_polarization_layer_separation}
\end{figure*}

As an example, consider the problem of obtaining the energy of the interlayer-exciton condensate state at $\nu{=}1$ in a bilayer system~\cite{Eisenstein14}. We model the interlayer-exciton condensate by the Halperin $(H)-(1,1,1)$ state~\cite{Halperin83}, described by the wavefunction
\begin{eqnarray}
\label{eq: Halperin_mmpn_wavefunction}
    \Psi^{H-(m,m',n)}_{\nu{=}\frac{m+m'-2n}{m m'-n^2}}&{=}&
    \prod_{1{\leq}i{<}j{\leq}N_{\uparrow}}(z^{\uparrow}_{i}{-}z^{\uparrow}_{j})^{m}
    \prod_{1{\leq}i{<}j{\leq}N_{\downarrow}}(z^{\downarrow}_{i}{-}z^{\downarrow}_{j})^{m'} \nonumber \\
    &&\prod_{\substack{1{\leq}i{\leq}N_{\uparrow} \\1{\leq}j{\leq}N_{\downarrow}}}(z^{\uparrow}_{i}{-}z^{\downarrow}_{j})^{n}e^{{-}\sum\limits_{1{\leq}i{\leq}N} \frac{|z_i|^{2}}{4\ell^2}}, 
\end{eqnarray}
where $m{=}m'{=}n{=}1$. Here, $z^{\sigma}_{l}$ is the two-dimensional coordinate of the $l^{\rm th}$ electron with spin-$\sigma$ parameterized as a complex number. For the Halperin-$(1,1,1)$ state with polarization $\gamma$, the various pair-correlation functions are
\begin{eqnarray}
\label{eq: pair_correlation_nu_1_Halperin_111_disk}
    g^{\nu{=}1~H-(1,1,1)\uparrow,\uparrow}\left(r\right)&{=}& \left(\frac{1+\gamma}{2}\right)^{2} \left[1{-}e^{-\frac{r^2}{2\ell^2}} \right],  \nonumber \\
    g^{\nu{=}1~H-(1,1,1)\downarrow,\downarrow}\left(r\right)&{=}& \left(\frac{1-\gamma}{2}\right)^{2} \left[1{-}e^{-\frac{r^2}{2\ell^2}} \right],  \nonumber \\
    g^{\nu{=}1~H-(1,1,1)\uparrow,\downarrow}\left(r\right)&{=}&g^{\nu{=}1~H-(1,1,1)\downarrow,\uparrow}\left(r\right) \\
    &{=}&\left(\frac{1+\gamma}{2}\right)\left(\frac{1-\gamma}{2}\right) \left[1{-}e^{-\frac{r^2}{2\ell^2}} \right].\nonumber
\end{eqnarray}
To obtain Eq.~\eqref{eq: pair_correlation_nu_1_Halperin_111_disk}, we made use of Eq.~\eqref{eq: pair_correlation_Sz_multiplet_thermodynamic}, along with the well-known pair-correlation function of the $\nu{=}1$ integer quantum Hall (IQH) state, which is $g^{\nu=1}(r){=}1{-}e^{-r^2/(2\ell^2)}$~\cite{Giuliani08}. Using these, the exact energy obtained from Eq.~\eqref{eq: energy_background_subtracted_per_particle_pair_correlation_thermodynamic} is
\begin{equation}
    \label{eq: energy_Halperin_111_finite_d}
    \frac{E^{\nu{=}1~H{-}(1,1,1)}_{0}}{e^{2}/(\varepsilon\ell)}{=}\sqrt{\frac{\pi }{32}} \left[{-}\gamma ^2{-}1{+}\left(\gamma ^2{-}1\right) e^{\frac{d^2}{2\ell^2}} \text{erfc}\left(\frac{d}{\sqrt{2}\ell}\right)\right],
\end{equation}
where $\text{erfc}(x)$ is the complementary error function, defined as $\text{erfc}(x){=}(2/\sqrt{\pi})\int_{x}^{\infty}e^{-t^2}dt$. For zero layer separation, i.e., $d{=}0$, and arbitrary $\gamma$, as well as for full polarizion in one layer, i.e., $\gamma{=}1$, and arbitrary $d$, $E^{\nu{=}1~H-(1,1,1)}_{0}{=}{-}\sqrt{\pi/8}~e^{2}/(\varepsilon\ell)$, which is the well-known energy of the fully polarized $\nu{=}1$ IQH state. A density plot of this energy of the Halperin-$(1,1,1)$ state, as a function of $d$ and $\gamma$, is shown in Fig.~\ref{fig: Coulomb_energy_FQH_states_polarization_layer_separation}$(a)$. 

Next, consider the $\nu{=}1/3$ Halperin-$(3,3,3)$ state. Using Eq.~\eqref{eq: pair_correlation_Sz_multiplet_thermodynamic}, we can get its pair-correlation function from that of the $1/3$ Laughlin~\cite{Laughlin83} pair-correlation function. For the $g(r)$ of the $1/3$ Laughlin state, we use an approximate form that has been obtained previously in the thermodynamic limit by fitting a particular functional form to finite-size Monte Carlo data~\cite{Girvin86, Dora24}. From the pair-correlation functions of the Halperin-$(3,3,3)$ state, we can obtain its energy using Eq.~\eqref{eq: energy_background_subtracted_per_particle_pair_correlation_thermodynamic}. For other fully polarized FQH states, one can proceed similarly. For the states at $2/3$ and $4/5$, their density-density correlators can be obtained from those of $1/3$ and $1/5$ Laughlin states, respectively, via particle-hole conjugation~\cite{Balram15b, Nguyen17}. Thus, we refer to the $2/3$ and $4/5$ states as anti-Laughlin. The pair-correlation function and static structure factor of many of the Laughlin~\cite{Laughlin83}, Jain~\cite{Jain89}, Moore-Read~\cite{Moore91}, etc., states have already been evaluated in Refs.~\cite{Girvin86, Fulsebakke23, Dora24}. We make use of these, along with the result of Eq.~\eqref{eq: pair_correlation_Sz_multiplet_thermodynamic}, to obtain the energies of these states for various $d$ and $\gamma$ using Eq.~\eqref{eq: energy_background_subtracted_per_particle_pair_correlation_thermodynamic}. A density plot of the energies of many FQH states, as a function of $d$ and $\gamma$, is shown in Fig.~\ref{fig: Coulomb_energy_FQH_states_polarization_layer_separation}. Note that the only error or uncertainty in the computed energies arises from the approximate form of the $g(r)$ being used. 

As the two layers get far-separated, i.e., as $d{\to}\infty$, the state at $\nu{=}1$ transitions to a product of two decoupled $1/2$ composite fermion Fermi seas~\cite{Halperin93, Rezayi94}, the energy of which is ${\approx}{-}0.466~e^{2}/(\varepsilon\ell)$~\cite{Balram15c, Balram17}. As $d{\to}\infty$, the states at $2/3$, $2/5$, $4/5$, and $4/9$, transition to a product of two decoupled $1/3$ Laughlin [equivalently, Halperin-$(3,3,0)$], $1/5$ Laughlin [equivalently, Halperin-$(5,5,0)$], $2/5$ Jain, and $2/9$ Jain states, respectively; the per-particle energies of which are ${\approx}{-}0.410~e^{2}/(\varepsilon\ell)$~\cite{Girvin86, Ciftja03}, ${\approx}{-}0.328~e^{2}/(\varepsilon\ell)$~\cite{Dora23}, ${\approx}{-}0.433~e^{2}/(\varepsilon\ell)$~\cite{Jain97b}, ${\approx}{-}0.343~e^{2}/(\varepsilon\ell)$~\cite{Balram15}, respectively. For $\gamma{=}1$, the state is fully polarized in one layer, and its energy is just its Coulomb energy since $d$ plays no role in this setting. Similarly, for $d{=}0$, the energies for all values of $\gamma$ are the same since the interaction is $SU(2)$ invariant in this limit. For this $SU(2)$ limit, the Coulomb energies of these states have been computed previously, and our results are fully consistent with them~\cite{Girvin86, Jain97, Jain07, Balram15a, Balram17, Balram20b, Dora23}. At $d{=}0$, at  $\nu{=}2/3,~2/5,~3/7$ and $4/9$, unpolarized spin-singlet or partially polarized states are more favorable since they have lower energies than the fully polarized states~\cite{Balram15a}. Nevertheless, the singlet states necessarily require $\gamma{=}0$, and thus can be destabilized by density imbalance, and the fully polarized states, which can readily accommodate any imbalance, could become relevant then. The energies shown in Fig.~\ref{fig: Coulomb_energy_FQH_states_polarization_layer_separation} can be utilized to construct the phase diagram as a function of density imbalance and layer separation~\cite{Moon95, Scarola01b, Moller08a, Papic09, Peterson15, Isobe17, Zhu16, Zhu17, Lian18a, Faugno20, Wagner21, Sharma23}, and can then be compared against experiments~\cite{Eisenstein14, Liu19, Li19, Shi22, Nguyen24a, Nguyen24}. 

\textbf{\textit{Discussion}.} 
We presented exact identities relating the density-density correlation functions of all the members of the maximal spin multiplet. Once the pair-correlation function and static structure factor of a single reference state in the multiplet are known, the corresponding quantities for every other member in the same multiplet are analytically obtainable via Eqs.~\eqref{eq: pair_correlation_Sz_multiplet_thermodynamic} and~\eqref{eq: structure_factor_Sz_multiplet_thermodynamic}. Thus, just the knowledge of the density-density correlator for a single member of the maximal spin multiplet yields the energy for arbitrary two-body interactions for all the multiplet members in a single shot, eliminating the need for independent computations for each of them. Furthermore, our results reveal $SU(2)$ spin-rotation symmetry-imposed constraints on density-density correlation functions for states within the maximal spin multiplet, such as how their vanishing properties, if any, are related, thereby providing a direct link between internal degrees of freedom and measurable density-density correlators. We demonstrated the power of our method by evaluating thermodynamic energies of many fractional quantum Hall states in a bilayer setting. In the End Matter, we have presented analogous identities relating the density-density correlators for finite systems in the spherical geometry in the quantum Hall regime. 

Our results are valid for both fermionic and bosonic systems and rely only on the states possessing the maximal $SU(2)$ spin or pseudospin (arising from valley, orbital, or layer degrees of freedom) quantum number, making them applicable well beyond the specific fractional quantum Hall setting that we applied them to. In particular, the relations we presented apply directly to the valley-polarized and intervalley-coherent states in moir\'e systems such as multilayer graphene~\cite{Christos22, Kim23, Mukherjee25, FQAH_Pentalayer_Graphene_Ju_2024} and transition metal dichalcogenides~\cite{Xie22, Lian23, FQAH_MoTe2_Xu_2023a, FQAH_MoTe2_Xu_2023b, FQAH_MoTe2_Mak_Shan_2023, FQAHE_MoTe2_Li_2023}, where excitingly zero-field versions of fractional quantum Hall states have also been realized recently, and where computing density-density correlation functions across the various polarization sectors is numerically challenging and demanding. 

Generalizing our results to systems with three or more components, where analogous relations would arise from the symmetry group $SU(n)$ for $n{>}2$, would serve as a natural and important direction for future work. Another interesting avenue worth exploring would be to expand our pair-distribution results to $k{-}$density correlators for $k{>}2$, such as for triplet-distribution functions, and use those to compute energies for $k$-body interactions. 

\textbf{\textit{Data availability}.} All the data used in this study are publicly available on the GitHub repository~\cite{Kundu_Balram_density_correlators_multiplet_repo}.

\textbf{\textit{Acknowledgments}.} 
ACB acknowledges useful discussions with Rakesh K. Dora. RK acknowledges support from the NPDF scheme of the ANRF under Grant No. PDF/2025/002176. Computational portions of this work were undertaken on the Nandadevi and Kamet supercomputers, maintained and supported by the Institute of Mathematical Sciences' High-Performance Computing Center. Some of the numerical calculations were verified using the DiagHam libraries~\cite{DiagHam}. This research was supported in part by the International Centre for Theoretical Sciences (ICTS) through ACB's participation in the program Generalised symmetries and anomalies in quantum phases of matter 2026 (code: ICTS/ GSYQM2026/01).

\clearpage

\begin{center}
\textbf{\large End Matter}
\end{center}

\section{Exact relations between the density-density correlators of states in the highest spin multiplet in the spherical geometry in the quantum Hall regime}
\label{sec: sphere}
In this section, we present some useful results for the spherical geometry, which is commonly used to carry out computations for quantum Hall systems. The projected static structure factor, i.e., the static structure factor projected to the lowest Landau level (LLL) that is appropriate for the high-field limit, in the spherical geometry~\cite{Haldane83} of a uniform state $\left|\Psi_{0}\right\rangle$ is defined as~\cite{Simon94a, He94, Dora24, Dora25}
\begin{equation}
\label{eq: definition_projected_S(L)}
   \bar{S}^{\sigma, \sigma'}\left(L\right)=\frac{4\pi}{N}\left\langle\Psi_{0}\right|\left[\bar{\rho}^{~\sigma}_{L, M}\right]^{\dagger}\bar{\rho}_{L, M}^{~\sigma'}\left|\Psi_{0}\right\rangle,
\end{equation}
where $\sigma,{\sigma'}{=}\uparrow,\downarrow$, $\bar{\rho}_{L, M}^{~\sigma}$ is the projected density operator~\cite{Girvin85, Girvin86, Simon94a, He94, Dora24, Dora25} for the $\sigma{-}$component with orbital angular momentum quantum number $L$ and its $z{-}$component $M$, $2Q$ is the flux piercing the sphere, and $N$ is the total number of particles. The projected and unprojected static structure factors are related as~\cite{Dora24, Dora25}
\begin{eqnarray}
\label{eq: relation_unprojected_projected_structure_factor_fully_polarized_states}
 \bar{S}^{\sigma, \sigma'}\left(L\right)&=& S^{\sigma, \sigma'}\left(L\right)  \\
 &&-\delta_{\sigma, \sigma'}\frac{N_{\sigma}}{N}\left[1-(2Q+1)\left(\begin{array}{ccc}
Q & Q & L \\
-Q & Q & 0
\end{array}\right)^{2}\right],\nonumber
\end{eqnarray}
where the quantity in the circular brackets $()$ is the Wigner $3j$-symbol. The static structure factors are normalized such that $\bar{S}^{\sigma, \sigma'}\left(0\right){=} S^{\sigma, \sigma'}\left(0\right){=}N_{\sigma}N_{\sigma'}/N$.

The pair-correlation function, $g^{\sigma, \sigma'}\left(\theta\right)$, is related to the projected static structure factor $\bar{S}^{\sigma, \sigma'}\left(L\right)$ as~\cite{Dora25}
\begin{align}
\label{eq: pair_correlation_projected_structure_factor_relation}
    g^{\sigma, \sigma'}\left(\theta\right)&=\frac{1}{N}\sum_{L{=}0}^{2Q}(2L+1) P_{L}\left(\cos\left(\theta\right)\right) \\ 
    &\times \left[\bar{S}^{\sigma, \sigma'}\left(L\right)-\delta_{\sigma, \sigma'}\frac{N_{\sigma}}{N}(2Q+1)\left(\begin{array}{ccc}
Q & Q & L \\
-Q & Q & 0
\end{array}\right)^{2}\right], \nonumber
\end{align}
where $0{\leq}\theta{\leq}\pi$ denotes the angular distance on the sphere. The commonly used arc and chord distances on the sphere are related to $\theta$ as $r_{a}{=}\sqrt{Q}\theta\ell$ and $r_{c}{=}2\sqrt{Q}\sin(\theta/2)\ell$, respectively, where $\sqrt{Q}\ell$ is the radius of the sphere. For states filling the LLL, the projected static structure factor admits an analytic expression given by $\bar{S}^{\sigma, \sigma'}\left(L\right){=}\delta_{L,0}N_{\sigma}N_{\sigma'}/N$. Thus, the pair-correlation function for these filled-LLL states is~\cite{Dora25}
\begin{equation}
 \label{eq: pair_correlation_Slater_dets}
    g^{{\rm filled-LLL~}\sigma, \sigma'}\left(\theta\right)=
    \frac{N_{\sigma}N_{\sigma'}}{N^{2}}-\delta_{\sigma, \sigma'}\frac{(2Q+1)N_{\sigma}}{N^{2}}\left( \cos\left( \frac{\theta}{2}\right) \right)^{4Q}.   
\end{equation}
In particular, the pair-correlation function of the $\nu{=}1$ integer quantum Hall (IQH) state is given by~\cite{Dora25}
\begin{equation}
  \label{eq: pair_correlation_nu_1}
    g^{\nu=1}\left(\theta\right)=1 - \left( \cos\left( \frac{\theta}{2}\right) \right)^{2(N-1)},  
\end{equation}
and that of the spin-singlet $\nu{=}2$ IQH [equivalently, Halperin-$(1,1,0)$ of Eq.~\eqref{eq: Halperin_mmpn_wavefunction}] state is given by
\begin{eqnarray}
\label{eq: pair_correlation_nu_2_singlet}
    g^{\nu{=}2~{\rm singlet~}\uparrow,\uparrow}\left(\theta\right)&{=}&g^{\nu{=}2~{\rm singlet~}\downarrow,\downarrow}\left(\theta\right) \nonumber \\
    &{=}&\frac{1}{4} \left[1{-}\left( \cos\left( \frac{\theta}{2}\right) \right)^{2\left(\frac{N}{2}{-}1\right)} \right], \nonumber \\
    g^{\nu{=}2~{\rm singlet~}\uparrow,\downarrow}\left(\theta\right)&{=}&g^{\nu{=}2~{\rm singlet~}\downarrow,\uparrow}\left(\theta\right){=}\frac{1}{4}.
\end{eqnarray}
The pair-correlation function of the $\nu{=}1$ Halperin-$(1,1,1)$ state on the sphere can now be obtained using Eqs.~\eqref{eq: pair_correlation_nu_1} and~\eqref{eq: pair_correlation_Sz_multiplet}, and is given by
\begin{eqnarray}
\label{eq: pair_correlation_nu_1_Halperin_111}
    g^{\nu{=}1~H-(1,1,1)\uparrow,\uparrow}\left(\theta\right)&{=}& \frac{N_{\uparrow}(N_{\uparrow}-1)}{N(N{-}1)} \left[1{-}\left( \cos\left( \frac{\theta}{2}\right) \right)^{2\left(N{-}1\right)} \right],  \nonumber \\
    g^{\nu{=}1~H-(1,1,1)\downarrow,\downarrow}\left(\theta\right)&{=}&\frac{N_{\downarrow}(N_{\downarrow}-1)}{N(N{-}1)} \left[1{-}\left( \cos\left( \frac{\theta}{2}\right) \right)^{2\left(N{-}1\right)} \right], \nonumber \\
    g^{\nu{=}1~H-(1,1,1)\uparrow,\downarrow}\left(\theta\right)&{=}&g^{\nu{=}1~H-(1,1,1)\downarrow,\uparrow}\left(\theta\right) \\
    &{=}&\frac{N_{\uparrow}N_{\downarrow}}{N(N{-}1)}\left[1{-}\left( \cos\left( \frac{\theta}{2}\right) \right)^{2\left(N{-}1\right)} \right].\nonumber
\end{eqnarray}
The static structure factor is related to the pair-correlation function as~\cite{Dora25}, 
\begin{eqnarray}
 \label{eq: pair_correlation_structure_factor}
    S^{\sigma, \sigma'}\left(L\right)&=&\frac{\delta_{\sigma, \sigma'}N_{\sigma}}{N}+\frac{N}{2}\int_{0}^{\pi} g^{\sigma, \sigma'}(\theta)P_{L}(\cos(\theta)) \sin(\theta) d\theta, \nonumber \\
    S^{\uparrow, \uparrow}\left(L\right)&=&\frac{N_{\uparrow}}{N}+\frac{N}{2}\int_{0}^{\pi} g^{\uparrow,\uparrow}(\theta)P_{L}(\cos(\theta)) \sin(\theta) d\theta, \nonumber \\ 
    S^{\downarrow, \downarrow}\left(L\right)&=&\frac{N_{\downarrow}}{N}+\frac{N}{2}\int_{0}^{\pi} g^{\downarrow,\downarrow}(\theta)P_{L}(\cos(\theta)) \sin(\theta) d\theta, \nonumber \\
    S^{\uparrow, \downarrow}\left(L\right)&=&\frac{N}{2}\int_{0}^{\pi} g^{\uparrow,\downarrow}(\theta)P_{L}(\cos(\theta)) \sin(\theta) d\theta.
\end{eqnarray}
One can check that the normalization $S^{\sigma, \sigma'}\left(0\right){=}N_{\sigma}N_{\sigma'}/N$ is rightly met by the above equations. Using Eqs.~\eqref{eq: pair_correlation_Sz_multiplet} and~\eqref{eq: pair_correlation_structure_factor}, given the static structure factor, $S(L)$, for a fully polarized state with $\mathbb{S}{=}\mathbb{S}_{z}{=}N/2$, we can get the static structure factor for any of its $\mathbb{S}{=}N/2,~\mathbb{S}_{z}{<}N/2$ multiplets as
\begin{eqnarray}
 \label{eq: structure_factor_Sz_multiplet}
    S^{\uparrow, \uparrow}\left(L\right)&{=}&\frac{N_{\uparrow}}{N}{+}\frac{N_{\uparrow}(N_{\uparrow}-1)}{N(N-1)}\left[S(L)-1\right], \nonumber \\
    S^{\downarrow, \downarrow}\left(L\right)&{=}&\frac{N_{\downarrow}}{N}{+}\frac{N_{\downarrow}(N_{\downarrow}-1)}{N(N-1)}\left[S(L)-1\right],\\
    S^{\uparrow, \downarrow}\left(L\right)&{=}&\frac{N_{\uparrow}N_{\downarrow}}{N(N-1)}\left[S(L)-1\right]=S^{\downarrow, \uparrow}\left(L\right).\nonumber
\end{eqnarray}
In particular, the static structure factor of the $\nu{=}1$ Halperin-$(1,1,1)$ state on the sphere is given by
\begin{eqnarray}
\label{eq: structure_factor_nu_1_Halperin_111} 
S^{\nu{=}1~H-(1,1,1)\uparrow, \uparrow}\left(L\right)&{=}&\frac{N_{\uparrow}}{N}{+}\frac{N_{\uparrow}(N_{\uparrow}-1)}{N(N-1)}\left[S^{\nu{=}1}(L)-1\right], \nonumber \\
    S^{\nu{=}1~H-(1,1,1)\downarrow, \downarrow}\left(L\right)&{=}&\frac{N_{\downarrow}}{N}{+}\frac{N_{\downarrow}(N_{\downarrow}-1)}{N(N-1)}\left[S^{\nu{=}1}(L)-1\right],\nonumber\\
    S^{\nu{=}1~H-(1,1,1)\uparrow, \downarrow}\left(L\right)&{=}&S^{\nu{=}1~H-(1,1,1)\downarrow, \uparrow}\left(L\right)  \\
    &=&\frac{N_{\uparrow}N_{\downarrow}}{N(N-1)}\left[S^{\nu{=}1}(L)-1\right],\nonumber
\end{eqnarray}
where~\cite{Dora24, Dora25}
\begin{equation}
    S^{\nu{=}1}{=} N \delta_{L,0}+1-\frac{N!(N-1)!}{(N-1-L)!(N+L)!},
\end{equation}
since $\bar{S}^{\nu{=}1}{=} N \delta_{L,0}$ and we used Eq.~\eqref{eq: relation_unprojected_projected_structure_factor_fully_polarized_states} to go from the projected to the unprojected static structure factor. 

In terms of the projected static structure factor of Eq.~\eqref{eq: definition_projected_S(L)}, the energy of the uniform state $|\Psi_{0}\rangle$ is given as~\cite{Dora25}
\begin{eqnarray}
\label{eq: energies_from_projected_S(L)}
    \left\langle\bar{H}^{\sigma, \sigma'}\right\rangle_{\Psi_{0}}&=&\frac{N}{2}\sum_{L{=}0}^{2Q}v_{L}^{\sigma, \sigma'}\left(2L+1\right)\Bigg[\bar{S}^{\sigma, \sigma'}\left(L\right) \nonumber \\
    &&-\delta_{\sigma, \sigma'}\frac{N_{\sigma}}{N}(2Q+1)\left(\begin{array}{ccc}
Q & Q & L \\
-Q & Q & 0
\end{array}\right)^{2}\Bigg],
\end{eqnarray}
where $v_{L}^{\sigma, \sigma'}$ is the $L^{\rm th}$ harmonic of the $(\sigma,\sigma')$ component of the interaction $v(r)$~\cite{Dora24}. Let us assume the particles of interest are negatively charged electrons. To incorporate the contribution of the positively charged background, we take the following steps. Since the density of up spins is $N(1{+}\gamma)/2$ and that of down spins is $N(1{-}\gamma)/2$, the background  contribution to the energy, $E_{\rm bg}$, which needs to be subtracted from the electron-electron contribution to obtain the total energy, is~\cite{Sharma23}
\begin{eqnarray}
    \label{eq: background_contribution}
        E_{\rm bg} &=& \frac{N^2}{2}\mathcal{C}, \\ \nonumber
        \mathcal{C} &=& \left[ \frac{(1+\gamma)}{2} \right]^{2}~\mathcal{C}^{\uparrow,\uparrow}+\left[ \frac{(1-\gamma)}{2} \right]^{2}~\mathcal{C}^{\downarrow, \downarrow} \\ \nonumber
        && +2\left[ \frac{(1+\gamma)}{2} \right]\left[ \frac{(1-\gamma)}{2} \right]~\mathcal{C}^{\uparrow, \downarrow},
\end{eqnarray}
where $\mathcal{C}^{\sigma, \sigma'}$ is the charging energy~\cite{Balram20b} for the $(\sigma, \sigma')$ part of the interaction. Note that this is also the $L{=}0$ contribution of the static structure factor to the energy, i.e., $E_{\rm bg}{=}(N/2)\sum_{\sigma, \sigma'}v_{0}^{\sigma, \sigma'}N_{\sigma}N_{\sigma'}/N{=}(1/2)\sum_{\sigma, \sigma'}v_{0}^{\sigma, \sigma'}N_{\sigma}N_{\sigma'}$. Thus, to get the background-subtracted energies, one can simply set $\bar{S}^{\sigma, \sigma'}\left(0\right){=}0~\forall \sigma,\sigma'$ and then evaluate the expression given in Eq.~\eqref{eq: energies_from_projected_S(L)}. 

For inter-layer uncorrelated states, like Halperin-$(3,3,0)$ or Halperin-$(5,5,0)$~\cite{Halperin83}, the background-subtracted per-particle energy, even with finite-layer separation, is the same as that of the single-component $1/3$ Laughlin or $1/5$ Laughlin~\cite{Laughlin83} energies. This is because the up-down part of the electron-electron contribution is exactly canceled by the background since there are no up-down correlations. On the sphere, the background contribution of Eq.~\eqref{eq: background_contribution}, for the interaction considered in the main text, i.e., $v^{\uparrow, \uparrow}(r){=}1/r{=}v^{\downarrow, \downarrow}(r)$, $v^{\uparrow, \downarrow}(r, d){=}1/\sqrt{r^2{+}d^2}{=}v^{\downarrow, \uparrow}(r, d)$, is
\begin{equation}
        E^{\rm sphere}_{\rm bg}{=}\frac{N_{\uparrow}^2}{2\sqrt{Q}}{+}\frac{N_{\downarrow}^2}{2\sqrt{Q}}{+}\frac{N_{\uparrow}N_{\downarrow}}{2Q}\left(\sqrt{4Q{+}d^2}{-}d\right)~\frac{e^2}{\varepsilon\ell}.
\end{equation}
The results presented in this section can be utilized to compute quantities for finite systems in the spherical geometry in the quantum Hall regime.

\bibliography{biblio_fqhe}

\begin{thebibliography}{75}%
\makeatletter
\providecommand \@ifxundefined [1]{%
 \@ifx{#1\undefined}
}%
\providecommand \@ifnum [1]{%
 \ifnum #1\expandafter \@firstoftwo
 \else \expandafter \@secondoftwo
 \fi
}%
\providecommand \@ifx [1]{%
 \ifx #1\expandafter \@firstoftwo
 \else \expandafter \@secondoftwo
 \fi
}%
\providecommand \natexlab [1]{#1}%
\providecommand \enquote  [1]{``#1''}%
\providecommand \bibnamefont  [1]{#1}%
\providecommand \bibfnamefont [1]{#1}%
\providecommand \citenamefont [1]{#1}%
\providecommand \href@noop [0]{\@secondoftwo}%
\providecommand \href [0]{\begingroup \@sanitize@url \@href}%
\providecommand \@href[1]{\@@startlink{#1}\@@href}%
\providecommand \@@href[1]{\endgroup#1\@@endlink}%
\providecommand \@sanitize@url [0]{\catcode `\\12\catcode `\$12\catcode `\&12\catcode `\#12\catcode `\^12\catcode `\_12\catcode `\%12\relax}%
\providecommand \@@startlink[1]{}%
\providecommand \@@endlink[0]{}%
\providecommand \url  [0]{\begingroup\@sanitize@url \@url }%
\providecommand \@url [1]{\endgroup\@href {#1}{\urlprefix }}%
\providecommand \urlprefix  [0]{URL }%
\providecommand \Eprint [0]{\href }%
\providecommand \doibase [0]{https://doi.org/}%
\providecommand \selectlanguage [0]{\@gobble}%
\providecommand \bibinfo  [0]{\@secondoftwo}%
\providecommand \bibfield  [0]{\@secondoftwo}%
\providecommand \translation [1]{[#1]}%
\providecommand \BibitemOpen [0]{}%
\providecommand \bibitemStop [0]{}%
\providecommand \bibitemNoStop [0]{.\EOS\space}%
\providecommand \EOS [0]{\spacefactor3000\relax}%
\providecommand \BibitemShut  [1]{\csname bibitem#1\endcsname}%
\let\auto@bib@innerbib\@empty
\bibitem [{\citenamefont {Kubo}(1957)}]{Kubo57}%
  \BibitemOpen
  \bibfield  {author} {\bibinfo {author} {\bibfnamefont {R.}~\bibnamefont {Kubo}},\ }\bibfield  {title} {\bibinfo {title} {{Statistical}-{Mechanical} {Theory} of {Irreversible} {Processes}. {I}. {General} {Theory} and {Simple} {Applications} to {Magnetic} and {Conduction} {Problems}},\ }\href {https://doi.org/10.1143/jpsj.12.570} {\bibfield  {journal} {\bibinfo  {journal} {Journal of the Physical Society of Japan}\ }\textbf {\bibinfo {volume} {12}},\ \bibinfo {pages} {570–586} (\bibinfo {year} {1957})}\BibitemShut {NoStop}%
\bibitem [{\citenamefont {Hansen}\ and\ \citenamefont {McDonald}(2013)}]{Hansen13}%
  \BibitemOpen
  \bibfield  {author} {\bibinfo {author} {\bibfnamefont {J.-P.}\ \bibnamefont {Hansen}}\ and\ \bibinfo {author} {\bibfnamefont {I.~R.}\ \bibnamefont {McDonald}},\ }\href@noop {} {\emph {\bibinfo {title} {Theory of simple liquids: with applications to soft matter}}}\ (\bibinfo  {publisher} {Academic press},\ \bibinfo {year} {2013})\BibitemShut {NoStop}%
\bibitem [{\citenamefont {Friedrich}\ \emph {et~al.}(1913)\citenamefont {Friedrich}, \citenamefont {Knipping},\ and\ \citenamefont {Laue}}]{Friedrich13}%
  \BibitemOpen
  \bibfield  {author} {\bibinfo {author} {\bibfnamefont {W.}~\bibnamefont {Friedrich}}, \bibinfo {author} {\bibfnamefont {P.}~\bibnamefont {Knipping}},\ and\ \bibinfo {author} {\bibfnamefont {M.}~\bibnamefont {Laue}},\ }\bibfield  {title} {\bibinfo {title} {Interferenzerscheinungen bei {R\"ontgenstrahlen}},\ }\href {https://doi.org/https://doi.org/10.1002/andp.19133461004} {\bibfield  {journal} {\bibinfo  {journal} {Annalen der Physik}\ }\textbf {\bibinfo {volume} {346}},\ \bibinfo {pages} {971} (\bibinfo {year} {1913})}\BibitemShut {NoStop}%
\bibitem [{\citenamefont {Franklin}\ and\ \citenamefont {Gosling}(1953)}]{Franklin53}%
  \BibitemOpen
  \bibfield  {author} {\bibinfo {author} {\bibfnamefont {R.~E.}\ \bibnamefont {Franklin}}\ and\ \bibinfo {author} {\bibfnamefont {R.~G.}\ \bibnamefont {Gosling}},\ }\bibfield  {title} {\bibinfo {title} {{Molecular} {Configuration} in {Sodium} {Thymonucleate}},\ }\href {https://doi.org/10.1038/171740a0} {\bibfield  {journal} {\bibinfo  {journal} {Nature}\ }\textbf {\bibinfo {volume} {171}},\ \bibinfo {pages} {740–741} (\bibinfo {year} {1953})}\BibitemShut {NoStop}%
\bibitem [{\citenamefont {Ashcroft}\ and\ \citenamefont {Lekner}(1966)}]{Ashcroft66}%
  \BibitemOpen
  \bibfield  {author} {\bibinfo {author} {\bibfnamefont {N.~W.}\ \bibnamefont {Ashcroft}}\ and\ \bibinfo {author} {\bibfnamefont {J.}~\bibnamefont {Lekner}},\ }\bibfield  {title} {\bibinfo {title} {{Structure} and {Resistivity} of {Liquid} {Metals}},\ }\href {https://doi.org/10.1103/physrev.145.83} {\bibfield  {journal} {\bibinfo  {journal} {Physical Review}\ }\textbf {\bibinfo {volume} {145}},\ \bibinfo {pages} {83–90} (\bibinfo {year} {1966})}\BibitemShut {NoStop}%
\bibitem [{\citenamefont {Greenfield}\ \emph {et~al.}(1971)\citenamefont {Greenfield}, \citenamefont {Wellendorf},\ and\ \citenamefont {Wiser}}]{Greenfield71}%
  \BibitemOpen
  \bibfield  {author} {\bibinfo {author} {\bibfnamefont {A.~J.}\ \bibnamefont {Greenfield}}, \bibinfo {author} {\bibfnamefont {J.}~\bibnamefont {Wellendorf}},\ and\ \bibinfo {author} {\bibfnamefont {N.}~\bibnamefont {Wiser}},\ }\bibfield  {title} {\bibinfo {title} {{X-Ray} {Determination} of the {Static} {Structure} {Factor} of {Liquid} {Na} and {K}},\ }\href {https://doi.org/10.1103/physreva.4.1607} {\bibfield  {journal} {\bibinfo  {journal} {Physical Review A}\ }\textbf {\bibinfo {volume} {4}},\ \bibinfo {pages} {1607–1616} (\bibinfo {year} {1971})}\BibitemShut {NoStop}%
\bibitem [{\citenamefont {Yarnell}\ \emph {et~al.}(1973)\citenamefont {Yarnell}, \citenamefont {Katz}, \citenamefont {Wenzel},\ and\ \citenamefont {Koenig}}]{Yarnell73}%
  \BibitemOpen
  \bibfield  {author} {\bibinfo {author} {\bibfnamefont {J.~L.}\ \bibnamefont {Yarnell}}, \bibinfo {author} {\bibfnamefont {M.~J.}\ \bibnamefont {Katz}}, \bibinfo {author} {\bibfnamefont {R.~G.}\ \bibnamefont {Wenzel}},\ and\ \bibinfo {author} {\bibfnamefont {S.~H.}\ \bibnamefont {Koenig}},\ }\bibfield  {title} {\bibinfo {title} {{Structure} {Factor} and {Radial} {Distribution} {Function} for {Liquid} {Argon} at 85 {$^\circ$K}},\ }\href {https://doi.org/10.1103/physreva.7.2130} {\bibfield  {journal} {\bibinfo  {journal} {Physical Review A}\ }\textbf {\bibinfo {volume} {7}},\ \bibinfo {pages} {2130–2144} (\bibinfo {year} {1973})}\BibitemShut {NoStop}%
\bibitem [{\citenamefont {Vahvaselkä}(1978)}]{Vahvaselk78}%
  \BibitemOpen
  \bibfield  {author} {\bibinfo {author} {\bibfnamefont {K.~S.}\ \bibnamefont {Vahvaselkä}},\ }\bibfield  {title} {\bibinfo {title} {{X-Ray} {Diffraction} {Analysis} {of} {Liquid} {Hg}, {Sn}, {Zn}, {Al} {and} {Cu}},\ }\href {https://doi.org/10.1088/0031-8949/18/4/005} {\bibfield  {journal} {\bibinfo  {journal} {Physica Scripta}\ }\textbf {\bibinfo {volume} {18}},\ \bibinfo {pages} {266–274} (\bibinfo {year} {1978})}\BibitemShut {NoStop}%
\bibitem [{\citenamefont {Svensson}\ \emph {et~al.}(1980)\citenamefont {Svensson}, \citenamefont {Sears}, \citenamefont {Woods},\ and\ \citenamefont {Martel}}]{Svensson80}%
  \BibitemOpen
  \bibfield  {author} {\bibinfo {author} {\bibfnamefont {E.~C.}\ \bibnamefont {Svensson}}, \bibinfo {author} {\bibfnamefont {V.~F.}\ \bibnamefont {Sears}}, \bibinfo {author} {\bibfnamefont {A.~D.~B.}\ \bibnamefont {Woods}},\ and\ \bibinfo {author} {\bibfnamefont {P.}~\bibnamefont {Martel}},\ }\bibfield  {title} {\bibinfo {title} {Neutron-diffraction study of the static structure factor and pair correlations in liquid {$^{4}\mathrm{He}$}},\ }\href {https://doi.org/10.1103/PhysRevB.21.3638} {\bibfield  {journal} {\bibinfo  {journal} {Phys. Rev. B}\ }\textbf {\bibinfo {volume} {21}},\ \bibinfo {pages} {3638} (\bibinfo {year} {1980})}\BibitemShut {NoStop}%
\bibitem [{\citenamefont {Furrer}\ \emph {et~al.}(2009)\citenamefont {Furrer}, \citenamefont {Mesot},\ and\ \citenamefont {Str{\"a}ssle}}]{Furrer09}%
  \BibitemOpen
  \bibfield  {author} {\bibinfo {author} {\bibfnamefont {A.}~\bibnamefont {Furrer}}, \bibinfo {author} {\bibfnamefont {J.~F.}\ \bibnamefont {Mesot}},\ and\ \bibinfo {author} {\bibfnamefont {T.}~\bibnamefont {Str{\"a}ssle}},\ }\href@noop {} {\emph {\bibinfo {title} {Neutron scattering in condensed matter physics}}},\ Vol.~\bibinfo {volume} {4}\ (\bibinfo  {publisher} {World Scientific Publishing Company},\ \bibinfo {year} {2009})\BibitemShut {NoStop}%
\bibitem [{\citenamefont {Giuliani}\ and\ \citenamefont {Vignale}(2008)}]{Giuliani08}%
  \BibitemOpen
  \bibfield  {author} {\bibinfo {author} {\bibfnamefont {G.}~\bibnamefont {Giuliani}}\ and\ \bibinfo {author} {\bibfnamefont {G.}~\bibnamefont {Vignale}},\ }\href@noop {} {\emph {\bibinfo {title} {Quantum Theory of the Electron Liquid}}}\ (\bibinfo  {publisher} {Cambridge University Press, The Edinburgh Building, Cambridge CB2 2RU, UK},\ \bibinfo {year} {2008})\BibitemShut {NoStop}%
\bibitem [{\citenamefont {Mahan}(2013)}]{Mahan13}%
  \BibitemOpen
  \bibfield  {author} {\bibinfo {author} {\bibfnamefont {G.~D.}\ \bibnamefont {Mahan}},\ }\href@noop {} {\emph {\bibinfo {title} {Many-particle physics}}}\ (\bibinfo  {publisher} {Springer Science \& Business Media},\ \bibinfo {year} {2013})\BibitemShut {NoStop}%
\bibitem [{\citenamefont {Pines}(2018)}]{Pines18}%
  \BibitemOpen
  \bibfield  {author} {\bibinfo {author} {\bibfnamefont {D.}~\bibnamefont {Pines}},\ }\href@noop {} {\emph {\bibinfo {title} {Theory of quantum liquids: normal {Fermi} liquids}}}\ (\bibinfo  {publisher} {CRC Press},\ \bibinfo {year} {2018})\BibitemShut {NoStop}%
\bibitem [{\citenamefont {Yoshioka}\ \emph {et~al.}(1983)\citenamefont {Yoshioka}, \citenamefont {Halperin},\ and\ \citenamefont {Lee}}]{Yoshioka83}%
  \BibitemOpen
  \bibfield  {author} {\bibinfo {author} {\bibfnamefont {D.}~\bibnamefont {Yoshioka}}, \bibinfo {author} {\bibfnamefont {B.~I.}\ \bibnamefont {Halperin}},\ and\ \bibinfo {author} {\bibfnamefont {P.~A.}\ \bibnamefont {Lee}},\ }\bibfield  {title} {\bibinfo {title} {Ground state of two-dimensional electrons in strong magnetic fields and $\frac{1}{3}$ quantized {Hall} effect},\ }\href {https://doi.org/10.1103/PhysRevLett.50.1219} {\bibfield  {journal} {\bibinfo  {journal} {Phys. Rev. Lett.}\ }\textbf {\bibinfo {volume} {50}},\ \bibinfo {pages} {1219} (\bibinfo {year} {1983})}\BibitemShut {NoStop}%
\bibitem [{\citenamefont {Lin}(1990)}]{Lin90}%
  \BibitemOpen
  \bibfield  {author} {\bibinfo {author} {\bibfnamefont {H.}~\bibnamefont {Lin}},\ }\bibfield  {title} {\bibinfo {title} {Exact diagonalization of quantum-spin models},\ }\href@noop {} {\bibfield  {journal} {\bibinfo  {journal} {Physical Review B}\ }\textbf {\bibinfo {volume} {42}},\ \bibinfo {pages} {6561} (\bibinfo {year} {1990})}\BibitemShut {NoStop}%
\bibitem [{\citenamefont {Pereira}\ \emph {et~al.}(2007)\citenamefont {Pereira}, \citenamefont {Sirker}, \citenamefont {Caux}, \citenamefont {Hagemans}, \citenamefont {Maillet}, \citenamefont {White},\ and\ \citenamefont {Affleck}}]{Pereira07}%
  \BibitemOpen
  \bibfield  {author} {\bibinfo {author} {\bibfnamefont {R.}~\bibnamefont {Pereira}}, \bibinfo {author} {\bibfnamefont {J.}~\bibnamefont {Sirker}}, \bibinfo {author} {\bibfnamefont {J.}~\bibnamefont {Caux}}, \bibinfo {author} {\bibfnamefont {R.}~\bibnamefont {Hagemans}}, \bibinfo {author} {\bibfnamefont {J.~M.}\ \bibnamefont {Maillet}}, \bibinfo {author} {\bibfnamefont {S.}~\bibnamefont {White}},\ and\ \bibinfo {author} {\bibfnamefont {I.}~\bibnamefont {Affleck}},\ }\bibfield  {title} {\bibinfo {title} {Dynamical structure factor at small q for the xxz spin-1/2 chain},\ }\href@noop {} {\bibfield  {journal} {\bibinfo  {journal} {Journal of Statistical Mechanics: Theory and Experiment}\ }\textbf {\bibinfo {volume} {2007}},\ \bibinfo {pages} {P08022} (\bibinfo {year} {2007})}\BibitemShut {NoStop}%
\bibitem [{\citenamefont {Fehske}\ \emph {et~al.}(2007)\citenamefont {Fehske}, \citenamefont {Schneider},\ and\ \citenamefont {Weisse}}]{Fehske07}%
  \BibitemOpen
  \bibfield  {author} {\bibinfo {author} {\bibfnamefont {H.}~\bibnamefont {Fehske}}, \bibinfo {author} {\bibfnamefont {R.}~\bibnamefont {Schneider}},\ and\ \bibinfo {author} {\bibfnamefont {A.}~\bibnamefont {Weisse}},\ }\href@noop {} {\emph {\bibinfo {title} {Computational many-particle physics}}}\ (\bibinfo  {publisher} {Springer},\ \bibinfo {year} {2007})\BibitemShut {NoStop}%
\bibitem [{\citenamefont {Zhang}\ and\ \citenamefont {Dong}(2010)}]{Zhang10}%
  \BibitemOpen
  \bibfield  {author} {\bibinfo {author} {\bibfnamefont {J.}~\bibnamefont {Zhang}}\ and\ \bibinfo {author} {\bibfnamefont {R.}~\bibnamefont {Dong}},\ }\bibfield  {title} {\bibinfo {title} {Exact diagonalization: the {Bose}--{Hubbard} model as an example},\ }\href@noop {} {\bibfield  {journal} {\bibinfo  {journal} {European Journal of Physics}\ }\textbf {\bibinfo {volume} {31}},\ \bibinfo {pages} {591} (\bibinfo {year} {2010})}\BibitemShut {NoStop}%
\bibitem [{\citenamefont {Binder}\ and\ \citenamefont {Heermann}(2010)}]{Binder10}%
  \BibitemOpen
  \bibfield  {author} {\bibinfo {author} {\bibfnamefont {K.}~\bibnamefont {Binder}}\ and\ \bibinfo {author} {\bibfnamefont {D.}~\bibnamefont {Heermann}},\ }\href@noop {} {\emph {\bibinfo {title} {{Monte} {Carlo} Simulation in Statistical Physics}}}\ (\bibinfo  {publisher} {Springer-Verlag Berlin Heidelberg},\ \bibinfo {year} {2010})\BibitemShut {NoStop}%
\bibitem [{\citenamefont {Dora}\ and\ \citenamefont {Balram}(2026)}]{Dora25}%
  \BibitemOpen
  \bibfield  {author} {\bibinfo {author} {\bibfnamefont {R.~K.}\ \bibnamefont {Dora}}\ and\ \bibinfo {author} {\bibfnamefont {A.~C.}\ \bibnamefont {Balram}},\ }\bibfield  {title} {\bibinfo {title} {Dispersion of collective modes in spinful fractional quantum {Hall} states on the sphere},\ }\href {https://doi.org/10.1103/17p5-jkym} {\bibfield  {journal} {\bibinfo  {journal} {Phys. Rev. B}\ }\textbf {\bibinfo {volume} {113}},\ \bibinfo {pages} {115420} (\bibinfo {year} {2026})}\BibitemShut {NoStop}%
\bibitem [{\citenamefont {Eisenstein}(2014)}]{Eisenstein14}%
  \BibitemOpen
  \bibfield  {author} {\bibinfo {author} {\bibfnamefont {J.}~\bibnamefont {Eisenstein}},\ }\bibfield  {title} {\bibinfo {title} {Exciton condensation in bilayer quantum {Hall} systems},\ }\href@noop {} {\bibfield  {journal} {\bibinfo  {journal} {Annu. Rev. Condens. Matter Phys.}\ }\textbf {\bibinfo {volume} {5}},\ \bibinfo {pages} {159} (\bibinfo {year} {2014})}\BibitemShut {NoStop}%
\bibitem [{\citenamefont {Halperin}(1983)}]{Halperin83}%
  \BibitemOpen
  \bibfield  {author} {\bibinfo {author} {\bibfnamefont {B.~I.}\ \bibnamefont {Halperin}},\ }\bibfield  {title} {\bibinfo {title} {Theory of the quantized {Hall} conductance},\ }\href@noop {} {\bibfield  {journal} {\bibinfo  {journal} {Helvetica Physica Acta}\ }\textbf {\bibinfo {volume} {56}},\ \bibinfo {pages} {75} (\bibinfo {year} {1983})}\BibitemShut {NoStop}%
\bibitem [{\citenamefont {Laughlin}(1983)}]{Laughlin83}%
  \BibitemOpen
  \bibfield  {author} {\bibinfo {author} {\bibfnamefont {R.~B.}\ \bibnamefont {Laughlin}},\ }\bibfield  {title} {\bibinfo {title} {Anomalous quantum {Hall} effect: An incompressible quantum fluid with fractionally charged excitations},\ }\href {https://doi.org/10.1103/PhysRevLett.50.1395} {\bibfield  {journal} {\bibinfo  {journal} {Phys. Rev. Lett.}\ }\textbf {\bibinfo {volume} {50}},\ \bibinfo {pages} {1395} (\bibinfo {year} {1983})}\BibitemShut {NoStop}%
\bibitem [{\citenamefont {Girvin}\ \emph {et~al.}(1986)\citenamefont {Girvin}, \citenamefont {MacDonald},\ and\ \citenamefont {Platzman}}]{Girvin86}%
  \BibitemOpen
  \bibfield  {author} {\bibinfo {author} {\bibfnamefont {S.~M.}\ \bibnamefont {Girvin}}, \bibinfo {author} {\bibfnamefont {A.~H.}\ \bibnamefont {MacDonald}},\ and\ \bibinfo {author} {\bibfnamefont {P.~M.}\ \bibnamefont {Platzman}},\ }\bibfield  {title} {\bibinfo {title} {Magneto-roton theory of collective excitations in the fractional quantum {Hall} effect},\ }\href {https://doi.org/10.1103/PhysRevB.33.2481} {\bibfield  {journal} {\bibinfo  {journal} {Phys. Rev. B}\ }\textbf {\bibinfo {volume} {33}},\ \bibinfo {pages} {2481} (\bibinfo {year} {1986})}\BibitemShut {NoStop}%
\bibitem [{\citenamefont {Dora}\ and\ \citenamefont {Balram}(2025)}]{Dora24}%
  \BibitemOpen
  \bibfield  {author} {\bibinfo {author} {\bibfnamefont {R.~K.}\ \bibnamefont {Dora}}\ and\ \bibinfo {author} {\bibfnamefont {A.~C.}\ \bibnamefont {Balram}},\ }\bibfield  {title} {\bibinfo {title} {Static structure factor and the dispersion of the {Girvin}-{MacDonald}-{Platzman} density mode for fractional quantum {Hall} fluids on the {Haldane} sphere},\ }\href {https://doi.org/10.1103/PhysRevB.111.115132} {\bibfield  {journal} {\bibinfo  {journal} {Phys. Rev. B}\ }\textbf {\bibinfo {volume} {111}},\ \bibinfo {pages} {115132} (\bibinfo {year} {2025})}\BibitemShut {NoStop}%
\bibitem [{\citenamefont {Balram}\ \emph {et~al.}(2015{\natexlab{a}})\citenamefont {Balram}, \citenamefont {T\ifmmode~\mbox{\H{o}}\else \H{o}\fi{}ke},\ and\ \citenamefont {Jain}}]{Balram15b}%
  \BibitemOpen
  \bibfield  {author} {\bibinfo {author} {\bibfnamefont {A.~C.}\ \bibnamefont {Balram}}, \bibinfo {author} {\bibfnamefont {C.}~\bibnamefont {T\ifmmode~\mbox{\H{o}}\else \H{o}\fi{}ke}},\ and\ \bibinfo {author} {\bibfnamefont {J.~K.}\ \bibnamefont {Jain}},\ }\bibfield  {title} {\bibinfo {title} {Luttinger theorem for the strongly correlated {Fermi} liquid of composite fermions},\ }\href {https://doi.org/10.1103/PhysRevLett.115.186805} {\bibfield  {journal} {\bibinfo  {journal} {Phys. Rev. Lett.}\ }\textbf {\bibinfo {volume} {115}},\ \bibinfo {pages} {186805} (\bibinfo {year} {2015}{\natexlab{a}})}\BibitemShut {NoStop}%
\bibitem [{\citenamefont {Nguyen}\ \emph {et~al.}(2017)\citenamefont {Nguyen}, \citenamefont {Can},\ and\ \citenamefont {Gromov}}]{Nguyen17}%
  \BibitemOpen
  \bibfield  {author} {\bibinfo {author} {\bibfnamefont {D.~X.}\ \bibnamefont {Nguyen}}, \bibinfo {author} {\bibfnamefont {T.}~\bibnamefont {Can}},\ and\ \bibinfo {author} {\bibfnamefont {A.}~\bibnamefont {Gromov}},\ }\bibfield  {title} {\bibinfo {title} {Particle-hole duality in the lowest {Landau} level},\ }\href {https://doi.org/10.1103/PhysRevLett.118.206602} {\bibfield  {journal} {\bibinfo  {journal} {Phys. Rev. Lett.}\ }\textbf {\bibinfo {volume} {118}},\ \bibinfo {pages} {206602} (\bibinfo {year} {2017})}\BibitemShut {NoStop}%
\bibitem [{\citenamefont {Jain}(1989)}]{Jain89}%
  \BibitemOpen
  \bibfield  {author} {\bibinfo {author} {\bibfnamefont {J.~K.}\ \bibnamefont {Jain}},\ }\bibfield  {title} {\bibinfo {title} {Composite-fermion approach for the fractional quantum {Hall} effect},\ }\href {https://doi.org/10.1103/PhysRevLett.63.199} {\bibfield  {journal} {\bibinfo  {journal} {Phys. Rev. Lett.}\ }\textbf {\bibinfo {volume} {63}},\ \bibinfo {pages} {199} (\bibinfo {year} {1989})}\BibitemShut {NoStop}%
\bibitem [{\citenamefont {Moore}\ and\ \citenamefont {Read}(1991)}]{Moore91}%
  \BibitemOpen
  \bibfield  {author} {\bibinfo {author} {\bibfnamefont {G.}~\bibnamefont {Moore}}\ and\ \bibinfo {author} {\bibfnamefont {N.}~\bibnamefont {Read}},\ }\bibfield  {title} {\bibinfo {title} {Nonabelions in the fractional quantum {Hall} effect},\ }\href {https://doi.org/10.1016/0550-3213(91)90407-O} {\bibfield  {journal} {\bibinfo  {journal} {Nucl. Phys. B}\ }\textbf {\bibinfo {volume} {360}},\ \bibinfo {pages} {362 } (\bibinfo {year} {1991})}\BibitemShut {NoStop}%
\bibitem [{\citenamefont {Fulsebakke}\ \emph {et~al.}(2023)\citenamefont {Fulsebakke}, \citenamefont {Fremling}, \citenamefont {Moran},\ and\ \citenamefont {Slingerland}}]{Fulsebakke23}%
  \BibitemOpen
  \bibfield  {author} {\bibinfo {author} {\bibfnamefont {J.}~\bibnamefont {Fulsebakke}}, \bibinfo {author} {\bibfnamefont {M.}~\bibnamefont {Fremling}}, \bibinfo {author} {\bibfnamefont {N.}~\bibnamefont {Moran}},\ and\ \bibinfo {author} {\bibfnamefont {J.~K.}\ \bibnamefont {Slingerland}},\ }\bibfield  {title} {\bibinfo {title} {{Parametrization and thermodynamic scaling of pair correlation functions for the fractional quantum {Hall} effect}},\ }\href {https://doi.org/10.21468/SciPostPhys.14.6.149} {\bibfield  {journal} {\bibinfo  {journal} {SciPost Phys.}\ }\textbf {\bibinfo {volume} {14}},\ \bibinfo {pages} {149} (\bibinfo {year} {2023})}\BibitemShut {NoStop}%
\bibitem [{\citenamefont {Halperin}\ \emph {et~al.}(1993)\citenamefont {Halperin}, \citenamefont {Lee},\ and\ \citenamefont {Read}}]{Halperin93}%
  \BibitemOpen
  \bibfield  {author} {\bibinfo {author} {\bibfnamefont {B.~I.}\ \bibnamefont {Halperin}}, \bibinfo {author} {\bibfnamefont {P.~A.}\ \bibnamefont {Lee}},\ and\ \bibinfo {author} {\bibfnamefont {N.}~\bibnamefont {Read}},\ }\bibfield  {title} {\bibinfo {title} {Theory of the half-filled {Landau} level},\ }\href {https://doi.org/10.1103/PhysRevB.47.7312} {\bibfield  {journal} {\bibinfo  {journal} {Phys. Rev. B}\ }\textbf {\bibinfo {volume} {47}},\ \bibinfo {pages} {7312} (\bibinfo {year} {1993})}\BibitemShut {NoStop}%
\bibitem [{\citenamefont {Rezayi}\ and\ \citenamefont {Read}(1994)}]{Rezayi94}%
  \BibitemOpen
  \bibfield  {author} {\bibinfo {author} {\bibfnamefont {E.}~\bibnamefont {Rezayi}}\ and\ \bibinfo {author} {\bibfnamefont {N.}~\bibnamefont {Read}},\ }\bibfield  {title} {\bibinfo {title} {Fermi-liquid-like state in a half-filled {Landau} level},\ }\href {https://doi.org/10.1103/PhysRevLett.72.900} {\bibfield  {journal} {\bibinfo  {journal} {Phys. Rev. Lett.}\ }\textbf {\bibinfo {volume} {72}},\ \bibinfo {pages} {900} (\bibinfo {year} {1994})}\BibitemShut {NoStop}%
\bibitem [{\citenamefont {Balram}\ \emph {et~al.}(2015{\natexlab{b}})\citenamefont {Balram}, \citenamefont {T\ifmmode~\mbox{\H{o}}\else \H{o}\fi{}ke}, \citenamefont {W\'ojs},\ and\ \citenamefont {Jain}}]{Balram15c}%
  \BibitemOpen
  \bibfield  {author} {\bibinfo {author} {\bibfnamefont {A.~C.}\ \bibnamefont {Balram}}, \bibinfo {author} {\bibfnamefont {C.}~\bibnamefont {T\ifmmode~\mbox{\H{o}}\else \H{o}\fi{}ke}}, \bibinfo {author} {\bibfnamefont {A.}~\bibnamefont {W\'ojs}},\ and\ \bibinfo {author} {\bibfnamefont {J.~K.}\ \bibnamefont {Jain}},\ }\bibfield  {title} {\bibinfo {title} {Spontaneous polarization of composite fermions in the $n=1$ {Landau} level of graphene},\ }\href {https://doi.org/10.1103/PhysRevB.92.205120} {\bibfield  {journal} {\bibinfo  {journal} {Phys. Rev. B}\ }\textbf {\bibinfo {volume} {92}},\ \bibinfo {pages} {205120} (\bibinfo {year} {2015}{\natexlab{b}})}\BibitemShut {NoStop}%
\bibitem [{\citenamefont {Balram}\ and\ \citenamefont {Jain}(2017)}]{Balram17}%
  \BibitemOpen
  \bibfield  {author} {\bibinfo {author} {\bibfnamefont {A.~C.}\ \bibnamefont {Balram}}\ and\ \bibinfo {author} {\bibfnamefont {J.~K.}\ \bibnamefont {Jain}},\ }\bibfield  {title} {\bibinfo {title} {Fermi wave vector for the partially spin-polarized composite-fermion {Fermi} sea},\ }\href {https://doi.org/10.1103/PhysRevB.96.235102} {\bibfield  {journal} {\bibinfo  {journal} {Phys. Rev. B}\ }\textbf {\bibinfo {volume} {96}},\ \bibinfo {pages} {235102} (\bibinfo {year} {2017})}\BibitemShut {NoStop}%
\bibitem [{\citenamefont {Ciftja}\ and\ \citenamefont {Wexler}(2003)}]{Ciftja03}%
  \BibitemOpen
  \bibfield  {author} {\bibinfo {author} {\bibfnamefont {O.}~\bibnamefont {Ciftja}}\ and\ \bibinfo {author} {\bibfnamefont {C.}~\bibnamefont {Wexler}},\ }\bibfield  {title} {\bibinfo {title} {{Monte} {Carlo} simulation method for {Laughlin}-like states in a disk geometry},\ }\href {https://doi.org/10.1103/PhysRevB.67.075304} {\bibfield  {journal} {\bibinfo  {journal} {Phys. Rev. B}\ }\textbf {\bibinfo {volume} {67}},\ \bibinfo {pages} {075304} (\bibinfo {year} {2003})}\BibitemShut {NoStop}%
\bibitem [{\citenamefont {Dora}\ and\ \citenamefont {Balram}(2023)}]{Dora23}%
  \BibitemOpen
  \bibfield  {author} {\bibinfo {author} {\bibfnamefont {R.~K.}\ \bibnamefont {Dora}}\ and\ \bibinfo {author} {\bibfnamefont {A.~C.}\ \bibnamefont {Balram}},\ }\bibfield  {title} {\bibinfo {title} {Competition between fractional quantum {Hall} liquid and electron solid phases in the {Landau} levels of multilayer graphene},\ }\href {https://doi.org/10.1103/PhysRevB.108.235153} {\bibfield  {journal} {\bibinfo  {journal} {Phys. Rev. B}\ }\textbf {\bibinfo {volume} {108}},\ \bibinfo {pages} {235153} (\bibinfo {year} {2023})}\BibitemShut {NoStop}%
\bibitem [{\citenamefont {Jain}\ and\ \citenamefont {Kamilla}(1997{\natexlab{a}})}]{Jain97b}%
  \BibitemOpen
  \bibfield  {author} {\bibinfo {author} {\bibfnamefont {J.~K.}\ \bibnamefont {Jain}}\ and\ \bibinfo {author} {\bibfnamefont {R.~K.}\ \bibnamefont {Kamilla}},\ }\bibfield  {title} {\bibinfo {title} {Quantitative study of large composite-fermion systems},\ }\href {https://doi.org/10.1103/PhysRevB.55.R4895} {\bibfield  {journal} {\bibinfo  {journal} {Phys. Rev. B}\ }\textbf {\bibinfo {volume} {55}},\ \bibinfo {pages} {R4895} (\bibinfo {year} {1997}{\natexlab{a}})}\BibitemShut {NoStop}%
\bibitem [{\citenamefont {Balram}\ \emph {et~al.}(2015{\natexlab{c}})\citenamefont {Balram}, \citenamefont {T\"oke}, \citenamefont {W\'ojs},\ and\ \citenamefont {Jain}}]{Balram15}%
  \BibitemOpen
  \bibfield  {author} {\bibinfo {author} {\bibfnamefont {A.~C.}\ \bibnamefont {Balram}}, \bibinfo {author} {\bibfnamefont {C.}~\bibnamefont {T\"oke}}, \bibinfo {author} {\bibfnamefont {A.}~\bibnamefont {W\'ojs}},\ and\ \bibinfo {author} {\bibfnamefont {J.~K.}\ \bibnamefont {Jain}},\ }\bibfield  {title} {\bibinfo {title} {Phase diagram of fractional quantum {Hall} effect of composite fermions in multicomponent systems},\ }\href {https://doi.org/10.1103/PhysRevB.91.045109} {\bibfield  {journal} {\bibinfo  {journal} {Phys. Rev. B}\ }\textbf {\bibinfo {volume} {91}},\ \bibinfo {pages} {045109} (\bibinfo {year} {2015}{\natexlab{c}})}\BibitemShut {NoStop}%
\bibitem [{\citenamefont {Jain}\ and\ \citenamefont {Kamilla}(1997{\natexlab{b}})}]{Jain97}%
  \BibitemOpen
  \bibfield  {author} {\bibinfo {author} {\bibfnamefont {J.~K.}\ \bibnamefont {Jain}}\ and\ \bibinfo {author} {\bibfnamefont {R.~K.}\ \bibnamefont {Kamilla}},\ }\bibfield  {title} {\bibinfo {title} {Composite fermions in the {Hilbert} space of the lowest electronic {Landau} level},\ }\href {https://doi.org/10.1142/S0217979297001301} {\bibfield  {journal} {\bibinfo  {journal} {Int. J. Mod. Phys. B}\ }\textbf {\bibinfo {volume} {11}},\ \bibinfo {pages} {2621} (\bibinfo {year} {1997}{\natexlab{b}})}\BibitemShut {NoStop}%
\bibitem [{\citenamefont {Jain}(2007)}]{Jain07}%
  \BibitemOpen
  \bibfield  {author} {\bibinfo {author} {\bibfnamefont {J.~K.}\ \bibnamefont {Jain}},\ }\href@noop {} {\emph {\bibinfo {title} {Composite Fermions}}}\ (\bibinfo  {publisher} {Cambridge University Press, New York, US},\ \bibinfo {year} {2007})\BibitemShut {NoStop}%
\bibitem [{\citenamefont {Balram}\ \emph {et~al.}(2015{\natexlab{d}})\citenamefont {Balram}, \citenamefont {T\"oke}, \citenamefont {W\'ojs},\ and\ \citenamefont {Jain}}]{Balram15a}%
  \BibitemOpen
  \bibfield  {author} {\bibinfo {author} {\bibfnamefont {A.~C.}\ \bibnamefont {Balram}}, \bibinfo {author} {\bibfnamefont {C.}~\bibnamefont {T\"oke}}, \bibinfo {author} {\bibfnamefont {A.}~\bibnamefont {W\'ojs}},\ and\ \bibinfo {author} {\bibfnamefont {J.~K.}\ \bibnamefont {Jain}},\ }\bibfield  {title} {\bibinfo {title} {Fractional quantum {Hall} effect in graphene: Quantitative comparison between theory and experiment},\ }\href {https://doi.org/10.1103/PhysRevB.92.075410} {\bibfield  {journal} {\bibinfo  {journal} {Phys. Rev. B}\ }\textbf {\bibinfo {volume} {92}},\ \bibinfo {pages} {075410} (\bibinfo {year} {2015}{\natexlab{d}})}\BibitemShut {NoStop}%
\bibitem [{\citenamefont {Balram}\ and\ \citenamefont {W\'ojs}(2020)}]{Balram20b}%
  \BibitemOpen
  \bibfield  {author} {\bibinfo {author} {\bibfnamefont {A.~C.}\ \bibnamefont {Balram}}\ and\ \bibinfo {author} {\bibfnamefont {A.}~\bibnamefont {W\'ojs}},\ }\bibfield  {title} {\bibinfo {title} {Fractional quantum {Hall} effect at $\ensuremath{\nu}=2+4/9$},\ }\href {https://doi.org/10.1103/PhysRevResearch.2.032035} {\bibfield  {journal} {\bibinfo  {journal} {Phys. Rev. Research}\ }\textbf {\bibinfo {volume} {2}},\ \bibinfo {pages} {032035} (\bibinfo {year} {2020})}\BibitemShut {NoStop}%
\bibitem [{\citenamefont {Moon}\ \emph {et~al.}(1995)\citenamefont {Moon}, \citenamefont {Mori}, \citenamefont {Yang}, \citenamefont {Girvin}, \citenamefont {MacDonald}, \citenamefont {Zheng}, \citenamefont {Yoshioka},\ and\ \citenamefont {Zhang}}]{Moon95}%
  \BibitemOpen
  \bibfield  {author} {\bibinfo {author} {\bibfnamefont {K.}~\bibnamefont {Moon}}, \bibinfo {author} {\bibfnamefont {H.}~\bibnamefont {Mori}}, \bibinfo {author} {\bibfnamefont {K.}~\bibnamefont {Yang}}, \bibinfo {author} {\bibfnamefont {S.~M.}\ \bibnamefont {Girvin}}, \bibinfo {author} {\bibfnamefont {A.~H.}\ \bibnamefont {MacDonald}}, \bibinfo {author} {\bibfnamefont {L.}~\bibnamefont {Zheng}}, \bibinfo {author} {\bibfnamefont {D.}~\bibnamefont {Yoshioka}},\ and\ \bibinfo {author} {\bibfnamefont {S.-C.}\ \bibnamefont {Zhang}},\ }\bibfield  {title} {\bibinfo {title} {Spontaneous interlayer coherence in double-layer quantum {Hall} systems: Charged vortices and kosterlitz-thouless phase transitions},\ }\href {https://doi.org/10.1103/PhysRevB.51.5138} {\bibfield  {journal} {\bibinfo  {journal} {Phys. Rev. B}\ }\textbf {\bibinfo {volume} {95}},\ \bibinfo {pages} {5138} (\bibinfo {year} {1995})}\BibitemShut {NoStop}%
\bibitem [{\citenamefont {Scarola}\ and\ \citenamefont {Jain}(2001)}]{Scarola01b}%
  \BibitemOpen
  \bibfield  {author} {\bibinfo {author} {\bibfnamefont {V.~W.}\ \bibnamefont {Scarola}}\ and\ \bibinfo {author} {\bibfnamefont {J.~K.}\ \bibnamefont {Jain}},\ }\bibfield  {title} {\bibinfo {title} {Phase diagram of bilayer composite fermion states},\ }\href {https://doi.org/10.1103/PhysRevB.64.085313} {\bibfield  {journal} {\bibinfo  {journal} {Phys. Rev. B}\ }\textbf {\bibinfo {volume} {64}},\ \bibinfo {pages} {085313} (\bibinfo {year} {2001})}\BibitemShut {NoStop}%
\bibitem [{\citenamefont {M\"oller}\ \emph {et~al.}(2008)\citenamefont {M\"oller}, \citenamefont {Simon},\ and\ \citenamefont {Rezayi}}]{Moller08a}%
  \BibitemOpen
  \bibfield  {author} {\bibinfo {author} {\bibfnamefont {G.}~\bibnamefont {M\"oller}}, \bibinfo {author} {\bibfnamefont {S.~H.}\ \bibnamefont {Simon}},\ and\ \bibinfo {author} {\bibfnamefont {E.~H.}\ \bibnamefont {Rezayi}},\ }\bibfield  {title} {\bibinfo {title} {Paired composite fermion phase of quantum {Hall} bilayers at $\ensuremath{\nu}=\frac{1}{2}+\frac{1}{2}$},\ }\href {https://doi.org/10.1103/PhysRevLett.101.176803} {\bibfield  {journal} {\bibinfo  {journal} {Phys. Rev. Lett.}\ }\textbf {\bibinfo {volume} {101}},\ \bibinfo {pages} {176803} (\bibinfo {year} {2008})}\BibitemShut {NoStop}%
\bibitem [{\citenamefont {Papi\ifmmode~\acute{c}\else \'{c}\fi{}}\ \emph {et~al.}(2009)\citenamefont {Papi\ifmmode~\acute{c}\else \'{c}\fi{}}, \citenamefont {M\"oller}, \citenamefont {Milovanovi\ifmmode~\acute{c}\else \'{c}\fi{}}, \citenamefont {Regnault},\ and\ \citenamefont {Goerbig}}]{Papic09}%
  \BibitemOpen
  \bibfield  {author} {\bibinfo {author} {\bibfnamefont {Z.}~\bibnamefont {Papi\ifmmode~\acute{c}\else \'{c}\fi{}}}, \bibinfo {author} {\bibfnamefont {G.}~\bibnamefont {M\"oller}}, \bibinfo {author} {\bibfnamefont {M.~V.}\ \bibnamefont {Milovanovi\ifmmode~\acute{c}\else \'{c}\fi{}}}, \bibinfo {author} {\bibfnamefont {N.}~\bibnamefont {Regnault}},\ and\ \bibinfo {author} {\bibfnamefont {M.~O.}\ \bibnamefont {Goerbig}},\ }\bibfield  {title} {\bibinfo {title} {Fractional quantum {Hall} state at $\ensuremath{\nu}=\frac{1}{4}$ in a wide quantum well},\ }\href {https://doi.org/10.1103/PhysRevB.79.245325} {\bibfield  {journal} {\bibinfo  {journal} {Phys. Rev. B}\ }\textbf {\bibinfo {volume} {79}},\ \bibinfo {pages} {245325} (\bibinfo {year} {2009})}\BibitemShut {NoStop}%
\bibitem [{\citenamefont {Peterson}\ \emph {et~al.}(2015)\citenamefont {Peterson}, \citenamefont {Wu}, \citenamefont {Cheng}, \citenamefont {Barkeshli}, \citenamefont {Wang},\ and\ \citenamefont {Das~Sarma}}]{Peterson15}%
  \BibitemOpen
  \bibfield  {author} {\bibinfo {author} {\bibfnamefont {M.~R.}\ \bibnamefont {Peterson}}, \bibinfo {author} {\bibfnamefont {Y.-L.}\ \bibnamefont {Wu}}, \bibinfo {author} {\bibfnamefont {M.}~\bibnamefont {Cheng}}, \bibinfo {author} {\bibfnamefont {M.}~\bibnamefont {Barkeshli}}, \bibinfo {author} {\bibfnamefont {Z.}~\bibnamefont {Wang}},\ and\ \bibinfo {author} {\bibfnamefont {S.}~\bibnamefont {Das~Sarma}},\ }\bibfield  {title} {\bibinfo {title} {Abelian and non-{Abelian} states in $\ensuremath{\nu}=2/3$ bilayer fractional quantum {Hall} systems},\ }\href {https://doi.org/10.1103/PhysRevB.92.035103} {\bibfield  {journal} {\bibinfo  {journal} {Phys. Rev. B}\ }\textbf {\bibinfo {volume} {92}},\ \bibinfo {pages} {035103} (\bibinfo {year} {2015})}\BibitemShut {NoStop}%
\bibitem [{\citenamefont {Isobe}\ and\ \citenamefont {Fu}(2017)}]{Isobe17}%
  \BibitemOpen
  \bibfield  {author} {\bibinfo {author} {\bibfnamefont {H.}~\bibnamefont {Isobe}}\ and\ \bibinfo {author} {\bibfnamefont {L.}~\bibnamefont {Fu}},\ }\bibfield  {title} {\bibinfo {title} {Interlayer pairing symmetry of composite fermions in quantum {Hall} bilayers},\ }\href {https://doi.org/10.1103/PhysRevLett.118.166401} {\bibfield  {journal} {\bibinfo  {journal} {Phys. Rev. Lett.}\ }\textbf {\bibinfo {volume} {118}},\ \bibinfo {pages} {166401} (\bibinfo {year} {2017})}\BibitemShut {NoStop}%
\bibitem [{\citenamefont {Zhu}\ \emph {et~al.}(2016)\citenamefont {Zhu}, \citenamefont {Liu}, \citenamefont {Haldane},\ and\ \citenamefont {Sheng}}]{Zhu16}%
  \BibitemOpen
  \bibfield  {author} {\bibinfo {author} {\bibfnamefont {W.}~\bibnamefont {Zhu}}, \bibinfo {author} {\bibfnamefont {Z.}~\bibnamefont {Liu}}, \bibinfo {author} {\bibfnamefont {F.~D.~M.}\ \bibnamefont {Haldane}},\ and\ \bibinfo {author} {\bibfnamefont {D.~N.}\ \bibnamefont {Sheng}},\ }\bibfield  {title} {\bibinfo {title} {Fractional quantum {Hall} bilayers at half filling: Tunneling-driven non-abelian phase},\ }\href {https://doi.org/10.1103/PhysRevB.94.245147} {\bibfield  {journal} {\bibinfo  {journal} {Phys. Rev. B}\ }\textbf {\bibinfo {volume} {94}},\ \bibinfo {pages} {245147} (\bibinfo {year} {2016})}\BibitemShut {NoStop}%
\bibitem [{\citenamefont {Zhu}\ \emph {et~al.}(2017)\citenamefont {Zhu}, \citenamefont {Fu},\ and\ \citenamefont {Sheng}}]{Zhu17}%
  \BibitemOpen
  \bibfield  {author} {\bibinfo {author} {\bibfnamefont {Z.}~\bibnamefont {Zhu}}, \bibinfo {author} {\bibfnamefont {L.}~\bibnamefont {Fu}},\ and\ \bibinfo {author} {\bibfnamefont {D.~N.}\ \bibnamefont {Sheng}},\ }\bibfield  {title} {\bibinfo {title} {Numerical study of quantum {Hall} bilayers at total filling ${\ensuremath{\nu}}_{T}=1$: A new phase at intermediate layer distances},\ }\href {https://doi.org/10.1103/PhysRevLett.119.177601} {\bibfield  {journal} {\bibinfo  {journal} {Phys. Rev. Lett.}\ }\textbf {\bibinfo {volume} {119}},\ \bibinfo {pages} {177601} (\bibinfo {year} {2017})}\BibitemShut {NoStop}%
\bibitem [{\citenamefont {Lian}\ and\ \citenamefont {Zhang}(2018)}]{Lian18a}%
  \BibitemOpen
  \bibfield  {author} {\bibinfo {author} {\bibfnamefont {B.}~\bibnamefont {Lian}}\ and\ \bibinfo {author} {\bibfnamefont {S.-C.}\ \bibnamefont {Zhang}},\ }\bibfield  {title} {\bibinfo {title} {Wave function and emergent su(2) symmetry in the ${\ensuremath{\nu}}_{T}=1$ quantum {Hall} bilayer},\ }\href {https://doi.org/10.1103/PhysRevLett.120.077601} {\bibfield  {journal} {\bibinfo  {journal} {Phys. Rev. Lett.}\ }\textbf {\bibinfo {volume} {120}},\ \bibinfo {pages} {077601} (\bibinfo {year} {2018})}\BibitemShut {NoStop}%
\bibitem [{\citenamefont {Faugno}\ \emph {et~al.}(2020)\citenamefont {Faugno}, \citenamefont {Balram}, \citenamefont {W\'ojs},\ and\ \citenamefont {Jain}}]{Faugno20}%
  \BibitemOpen
  \bibfield  {author} {\bibinfo {author} {\bibfnamefont {W.~N.}\ \bibnamefont {Faugno}}, \bibinfo {author} {\bibfnamefont {A.~C.}\ \bibnamefont {Balram}}, \bibinfo {author} {\bibfnamefont {A.}~\bibnamefont {W\'ojs}},\ and\ \bibinfo {author} {\bibfnamefont {J.~K.}\ \bibnamefont {Jain}},\ }\bibfield  {title} {\bibinfo {title} {Theoretical phase diagram of two-component composite fermions in double-layer graphene},\ }\href {https://doi.org/10.1103/PhysRevB.101.085412} {\bibfield  {journal} {\bibinfo  {journal} {Phys. Rev. B}\ }\textbf {\bibinfo {volume} {101}},\ \bibinfo {pages} {085412} (\bibinfo {year} {2020})}\BibitemShut {NoStop}%
\bibitem [{\citenamefont {Wagner}\ \emph {et~al.}(2021)\citenamefont {Wagner}, \citenamefont {Nguyen}, \citenamefont {Simon},\ and\ \citenamefont {Halperin}}]{Wagner21}%
  \BibitemOpen
  \bibfield  {author} {\bibinfo {author} {\bibfnamefont {G.}~\bibnamefont {Wagner}}, \bibinfo {author} {\bibfnamefont {D.~X.}\ \bibnamefont {Nguyen}}, \bibinfo {author} {\bibfnamefont {S.~H.}\ \bibnamefont {Simon}},\ and\ \bibinfo {author} {\bibfnamefont {B.~I.}\ \bibnamefont {Halperin}},\ }\bibfield  {title} {\bibinfo {title} {$s$-wave paired electron and hole composite fermion trial state for quantum {Hall} bilayers with $\ensuremath{\nu}=1$},\ }\href {https://doi.org/10.1103/PhysRevLett.127.246803} {\bibfield  {journal} {\bibinfo  {journal} {Phys. Rev. Lett.}\ }\textbf {\bibinfo {volume} {127}},\ \bibinfo {pages} {246803} (\bibinfo {year} {2021})}\BibitemShut {NoStop}%
\bibitem [{\citenamefont {Sharma}\ \emph {et~al.}(2024)\citenamefont {Sharma}, \citenamefont {Balram},\ and\ \citenamefont {Jain}}]{Sharma23}%
  \BibitemOpen
  \bibfield  {author} {\bibinfo {author} {\bibfnamefont {A.}~\bibnamefont {Sharma}}, \bibinfo {author} {\bibfnamefont {A.~C.}\ \bibnamefont {Balram}},\ and\ \bibinfo {author} {\bibfnamefont {J.~K.}\ \bibnamefont {Jain}},\ }\bibfield  {title} {\bibinfo {title} {Composite-fermion pairing at half-filled and quarter-filled lowest {Landau} level},\ }\href {https://doi.org/10.1103/PhysRevB.109.035306} {\bibfield  {journal} {\bibinfo  {journal} {Phys. Rev. B}\ }\textbf {\bibinfo {volume} {109}},\ \bibinfo {pages} {035306} (\bibinfo {year} {2024})}\BibitemShut {NoStop}%
\bibitem [{\citenamefont {Liu}\ \emph {et~al.}(2019)\citenamefont {Liu}, \citenamefont {Hao}, \citenamefont {Watanabe}, \citenamefont {Taniguchi}, \citenamefont {Halperin},\ and\ \citenamefont {Kim}}]{Liu19}%
  \BibitemOpen
  \bibfield  {author} {\bibinfo {author} {\bibfnamefont {X.}~\bibnamefont {Liu}}, \bibinfo {author} {\bibfnamefont {Z.}~\bibnamefont {Hao}}, \bibinfo {author} {\bibfnamefont {K.}~\bibnamefont {Watanabe}}, \bibinfo {author} {\bibfnamefont {T.}~\bibnamefont {Taniguchi}}, \bibinfo {author} {\bibfnamefont {B.~I.}\ \bibnamefont {Halperin}},\ and\ \bibinfo {author} {\bibfnamefont {P.}~\bibnamefont {Kim}},\ }\bibfield  {title} {\bibinfo {title} {Interlayer fractional quantum {Hall} effect in a coupled graphene double layer},\ }\href {https://doi.org/10.1038/s41567-019-0546-0} {\bibfield  {journal} {\bibinfo  {journal} {Nature Physics}\ }\textbf {\bibinfo {volume} {15}},\ \bibinfo {pages} {893} (\bibinfo {year} {2019})}\BibitemShut {NoStop}%
\bibitem [{\citenamefont {Li}\ \emph {et~al.}(2019)\citenamefont {Li}, \citenamefont {Shi}, \citenamefont {Zeng}, \citenamefont {Watanabe}, \citenamefont {Taniguchi}, \citenamefont {Hone},\ and\ \citenamefont {Dean}}]{Li19}%
  \BibitemOpen
  \bibfield  {author} {\bibinfo {author} {\bibfnamefont {J.~I.~A.}\ \bibnamefont {Li}}, \bibinfo {author} {\bibfnamefont {Q.}~\bibnamefont {Shi}}, \bibinfo {author} {\bibfnamefont {Y.}~\bibnamefont {Zeng}}, \bibinfo {author} {\bibfnamefont {K.}~\bibnamefont {Watanabe}}, \bibinfo {author} {\bibfnamefont {T.}~\bibnamefont {Taniguchi}}, \bibinfo {author} {\bibfnamefont {J.}~\bibnamefont {Hone}},\ and\ \bibinfo {author} {\bibfnamefont {C.~R.}\ \bibnamefont {Dean}},\ }\bibfield  {title} {\bibinfo {title} {Pairing states of composite fermions in double-layer graphene},\ }\href {https://doi.org/10.1038/s41567-019-0547-z} {\bibfield  {journal} {\bibinfo  {journal} {Nature Physics}\ }\textbf {\bibinfo {volume} {15}},\ \bibinfo {pages} {898} (\bibinfo {year} {2019})}\BibitemShut {NoStop}%
\bibitem [{\citenamefont {Shi}\ \emph {et~al.}(2022)\citenamefont {Shi}, \citenamefont {Shih}, \citenamefont {Rhodes}, \citenamefont {Kim}, \citenamefont {Barmak}, \citenamefont {Watanabe}, \citenamefont {Taniguchi}, \citenamefont {Papi{\'c}}, \citenamefont {Abanin}, \citenamefont {Hone},\ and\ \citenamefont {Dean}}]{Shi22}%
  \BibitemOpen
  \bibfield  {author} {\bibinfo {author} {\bibfnamefont {Q.}~\bibnamefont {Shi}}, \bibinfo {author} {\bibfnamefont {E.-M.}\ \bibnamefont {Shih}}, \bibinfo {author} {\bibfnamefont {D.}~\bibnamefont {Rhodes}}, \bibinfo {author} {\bibfnamefont {B.}~\bibnamefont {Kim}}, \bibinfo {author} {\bibfnamefont {K.}~\bibnamefont {Barmak}}, \bibinfo {author} {\bibfnamefont {K.}~\bibnamefont {Watanabe}}, \bibinfo {author} {\bibfnamefont {T.}~\bibnamefont {Taniguchi}}, \bibinfo {author} {\bibfnamefont {Z.}~\bibnamefont {Papi{\'c}}}, \bibinfo {author} {\bibfnamefont {D.~A.}\ \bibnamefont {Abanin}}, \bibinfo {author} {\bibfnamefont {J.}~\bibnamefont {Hone}},\ and\ \bibinfo {author} {\bibfnamefont {C.~R.}\ \bibnamefont {Dean}},\ }\bibfield  {title} {\bibinfo {title} {Bilayer {W}{S}e2 as a natural platform for interlayer exciton condensates in the strong coupling limit},\ }\href {https://doi.org/10.1038/s41565-022-01104-5} {\bibfield  {journal} {\bibinfo  {journal} {Nature Nanotechnology}\ }\textbf {\bibinfo {volume} {17}},\ \bibinfo {pages} {577} (\bibinfo {year} {2022})}\BibitemShut {NoStop}%
\bibitem [{\citenamefont {Zhang}\ \emph {et~al.}(2025)\citenamefont {Zhang}, \citenamefont {Nguyen}, \citenamefont {Batra}, \citenamefont {Liu}, \citenamefont {Watanabe}, \citenamefont {Taniguchi}, \citenamefont {Feldman},\ and\ \citenamefont {Li}}]{Nguyen24a}%
  \BibitemOpen
  \bibfield  {author} {\bibinfo {author} {\bibfnamefont {N.~J.}\ \bibnamefont {Zhang}}, \bibinfo {author} {\bibfnamefont {R.~Q.}\ \bibnamefont {Nguyen}}, \bibinfo {author} {\bibfnamefont {N.}~\bibnamefont {Batra}}, \bibinfo {author} {\bibfnamefont {X.}~\bibnamefont {Liu}}, \bibinfo {author} {\bibfnamefont {K.}~\bibnamefont {Watanabe}}, \bibinfo {author} {\bibfnamefont {T.}~\bibnamefont {Taniguchi}}, \bibinfo {author} {\bibfnamefont {D.~E.}\ \bibnamefont {Feldman}},\ and\ \bibinfo {author} {\bibfnamefont {J.~I.~A.}\ \bibnamefont {Li}},\ }\bibfield  {title} {\bibinfo {title} {Excitons in the fractional quantum {Hall} effect},\ }\href {https://doi.org/10.1038/s41586-024-08274-3} {\bibfield  {journal} {\bibinfo  {journal} {Nature}\ }\textbf {\bibinfo {volume} {637}},\ \bibinfo {pages} {327} (\bibinfo {year} {2025})}\BibitemShut {NoStop}%
\bibitem [{\citenamefont {Nguyen}\ \emph {et~al.}(2024)\citenamefont {Nguyen}, \citenamefont {Zhang}, \citenamefont {Khurana-Batra}, \citenamefont {Alkidim}, \citenamefont {Liu}, \citenamefont {Watanabe}, \citenamefont {Taniguchi}, \citenamefont {Feldman},\ and\ \citenamefont {Li}}]{Nguyen24}%
  \BibitemOpen
  \bibfield  {author} {\bibinfo {author} {\bibfnamefont {R.~Q.}\ \bibnamefont {Nguyen}}, \bibinfo {author} {\bibfnamefont {N.~J.}\ \bibnamefont {Zhang}}, \bibinfo {author} {\bibfnamefont {N.}~\bibnamefont {Khurana-Batra}}, \bibinfo {author} {\bibfnamefont {S.}~\bibnamefont {Alkidim}}, \bibinfo {author} {\bibfnamefont {X.}~\bibnamefont {Liu}}, \bibinfo {author} {\bibfnamefont {K.}~\bibnamefont {Watanabe}}, \bibinfo {author} {\bibfnamefont {T.}~\bibnamefont {Taniguchi}}, \bibinfo {author} {\bibfnamefont {D.}~\bibnamefont {Feldman}},\ and\ \bibinfo {author} {\bibfnamefont {J.}~\bibnamefont {Li}},\ }\bibfield  {title} {\bibinfo {title} {Bilayer excitons in the laughlin fractional quantum {Hall} state},\ }\href@noop {} {\bibfield  {journal} {\bibinfo  {journal} {arXiv preprint arXiv:2410.24208}\ } (\bibinfo {year} {2024})}\BibitemShut {NoStop}%
\bibitem [{\citenamefont {Christos}\ \emph {et~al.}(2022)\citenamefont {Christos}, \citenamefont {Sachdev},\ and\ \citenamefont {Scheurer}}]{Christos22}%
  \BibitemOpen
  \bibfield  {author} {\bibinfo {author} {\bibfnamefont {M.}~\bibnamefont {Christos}}, \bibinfo {author} {\bibfnamefont {S.}~\bibnamefont {Sachdev}},\ and\ \bibinfo {author} {\bibfnamefont {M.~S.}\ \bibnamefont {Scheurer}},\ }\bibfield  {title} {\bibinfo {title} {Correlated insulators, semimetals, and superconductivity in twisted trilayer graphene},\ }\bibfield  {journal} {\bibinfo  {journal} {Physical Review X}\ }\textbf {\bibinfo {volume} {12}},\ \href {https://doi.org/10.1103/physrevx.12.021018} {10.1103/physrevx.12.021018} (\bibinfo {year} {2022})\BibitemShut {NoStop}%
\bibitem [{\citenamefont {Kim}\ \emph {et~al.}(2023)\citenamefont {Kim}, \citenamefont {Choi}, \citenamefont {Lantagne-Hurtubise}, \citenamefont {Lewandowski}, \citenamefont {Thomson}, \citenamefont {Kong}, \citenamefont {Zhou}, \citenamefont {Baum}, \citenamefont {Zhang}, \citenamefont {Holleis}, \citenamefont {Watanabe}, \citenamefont {Taniguchi}, \citenamefont {Young}, \citenamefont {Alicea},\ and\ \citenamefont {Nadj-Perge}}]{Kim23}%
  \BibitemOpen
  \bibfield  {author} {\bibinfo {author} {\bibfnamefont {H.}~\bibnamefont {Kim}}, \bibinfo {author} {\bibfnamefont {Y.}~\bibnamefont {Choi}}, \bibinfo {author} {\bibfnamefont {E.}~\bibnamefont {Lantagne-Hurtubise}}, \bibinfo {author} {\bibfnamefont {C.}~\bibnamefont {Lewandowski}}, \bibinfo {author} {\bibfnamefont {A.}~\bibnamefont {Thomson}}, \bibinfo {author} {\bibfnamefont {L.}~\bibnamefont {Kong}}, \bibinfo {author} {\bibfnamefont {H.}~\bibnamefont {Zhou}}, \bibinfo {author} {\bibfnamefont {E.}~\bibnamefont {Baum}}, \bibinfo {author} {\bibfnamefont {Y.}~\bibnamefont {Zhang}}, \bibinfo {author} {\bibfnamefont {L.}~\bibnamefont {Holleis}}, \bibinfo {author} {\bibfnamefont {K.}~\bibnamefont {Watanabe}}, \bibinfo {author} {\bibfnamefont {T.}~\bibnamefont {Taniguchi}}, \bibinfo {author} {\bibfnamefont {A.~F.}\ \bibnamefont {Young}}, \bibinfo {author} {\bibfnamefont {J.}~\bibnamefont {Alicea}},\ and\ \bibinfo {author} {\bibfnamefont {S.}~\bibnamefont {Nadj-Perge}},\ }\bibfield  {title} {\bibinfo {title} {Imaging inter-valley coherent order in magic-angle twisted trilayer graphene},\ }\href {https://doi.org/10.1038/s41586-023-06663-8} {\bibfield  {journal} {\bibinfo  {journal} {Nature}\ }\textbf {\bibinfo {volume} {623}},\ \bibinfo {pages} {942–948} (\bibinfo {year} {2023})}\BibitemShut {NoStop}%
\bibitem [{\citenamefont {Mukherjee}\ \emph {et~al.}(2025)\citenamefont {Mukherjee}, \citenamefont {Layek}, \citenamefont {Sinha}, \citenamefont {Kundu}, \citenamefont {Marchawala}, \citenamefont {Hingankar}, \citenamefont {Sarkar}, \citenamefont {Sangani}, \citenamefont {Agarwal}, \citenamefont {Ghosh}, \citenamefont {Tazi}, \citenamefont {Watanabe}, \citenamefont {Taniguchi}, \citenamefont {Pasupathy}, \citenamefont {Kundu},\ and\ \citenamefont {Deshmukh}}]{Mukherjee25}%
  \BibitemOpen
  \bibfield  {author} {\bibinfo {author} {\bibfnamefont {A.}~\bibnamefont {Mukherjee}}, \bibinfo {author} {\bibfnamefont {S.}~\bibnamefont {Layek}}, \bibinfo {author} {\bibfnamefont {S.}~\bibnamefont {Sinha}}, \bibinfo {author} {\bibfnamefont {R.}~\bibnamefont {Kundu}}, \bibinfo {author} {\bibfnamefont {A.~H.}\ \bibnamefont {Marchawala}}, \bibinfo {author} {\bibfnamefont {M.}~\bibnamefont {Hingankar}}, \bibinfo {author} {\bibfnamefont {J.}~\bibnamefont {Sarkar}}, \bibinfo {author} {\bibfnamefont {L.~D.~V.}\ \bibnamefont {Sangani}}, \bibinfo {author} {\bibfnamefont {H.}~\bibnamefont {Agarwal}}, \bibinfo {author} {\bibfnamefont {S.}~\bibnamefont {Ghosh}}, \bibinfo {author} {\bibfnamefont {A.~B.}\ \bibnamefont {Tazi}}, \bibinfo {author} {\bibfnamefont {K.}~\bibnamefont {Watanabe}}, \bibinfo {author} {\bibfnamefont {T.}~\bibnamefont {Taniguchi}}, \bibinfo {author} {\bibfnamefont {A.~N.}\ \bibnamefont {Pasupathy}}, \bibinfo {author} {\bibfnamefont {A.}~\bibnamefont {Kundu}},\ and\ \bibinfo {author} {\bibfnamefont {M.~M.}\ \bibnamefont {Deshmukh}},\ }\bibfield  {title} {\bibinfo {title} {Superconducting magic-angle twisted trilayer graphene with competing magnetic order and moiré inhomogeneities},\ }\href {https://doi.org/10.1038/s41563-025-02252-4} {\bibfield  {journal} {\bibinfo  {journal} {Nature Materials}\ }\textbf {\bibinfo {volume} {24}},\ \bibinfo {pages} {1400–1406} (\bibinfo {year} {2025})}\BibitemShut {NoStop}%
\bibitem [{\citenamefont {Lu}\ \emph {et~al.}(2024)\citenamefont {Lu}, \citenamefont {Han}, \citenamefont {Yao}, \citenamefont {Reddy}, \citenamefont {Yang}, \citenamefont {Seo}, \citenamefont {Watanabe}, \citenamefont {Taniguchi}, \citenamefont {Fu},\ and\ \citenamefont {Ju}}]{FQAH_Pentalayer_Graphene_Ju_2024}%
  \BibitemOpen
  \bibfield  {author} {\bibinfo {author} {\bibfnamefont {Z.}~\bibnamefont {Lu}}, \bibinfo {author} {\bibfnamefont {T.}~\bibnamefont {Han}}, \bibinfo {author} {\bibfnamefont {Y.}~\bibnamefont {Yao}}, \bibinfo {author} {\bibfnamefont {A.~P.}\ \bibnamefont {Reddy}}, \bibinfo {author} {\bibfnamefont {J.}~\bibnamefont {Yang}}, \bibinfo {author} {\bibfnamefont {J.}~\bibnamefont {Seo}}, \bibinfo {author} {\bibfnamefont {K.}~\bibnamefont {Watanabe}}, \bibinfo {author} {\bibfnamefont {T.}~\bibnamefont {Taniguchi}}, \bibinfo {author} {\bibfnamefont {L.}~\bibnamefont {Fu}},\ and\ \bibinfo {author} {\bibfnamefont {L.}~\bibnamefont {Ju}},\ }\bibfield  {title} {\bibinfo {title} {Fractional quantum anomalous {Hall} effect in multilayer graphene},\ }\href {https://doi.org/10.1038/s41586-023-07010-7} {\bibfield  {journal} {\bibinfo  {journal} {Nature}\ }\textbf {\bibinfo {volume} {626}},\ \bibinfo {pages} {759} (\bibinfo {year} {2024})}\BibitemShut {NoStop}%
\bibitem [{\citenamefont {Xie}\ \emph {et~al.}(2022)\citenamefont {Xie}, \citenamefont {Zhang}, \citenamefont {Hu}, \citenamefont {Mak},\ and\ \citenamefont {Law}}]{Xie22}%
  \BibitemOpen
  \bibfield  {author} {\bibinfo {author} {\bibfnamefont {Y.-M.}\ \bibnamefont {Xie}}, \bibinfo {author} {\bibfnamefont {C.-P.}\ \bibnamefont {Zhang}}, \bibinfo {author} {\bibfnamefont {J.-X.}\ \bibnamefont {Hu}}, \bibinfo {author} {\bibfnamefont {K.~F.}\ \bibnamefont {Mak}},\ and\ \bibinfo {author} {\bibfnamefont {K.~T.}\ \bibnamefont {Law}},\ }\bibfield  {title} {\bibinfo {title} {Valley-polarized quantum anomalous hall state in moir\'e ${\mathrm{mote}}_{2}/{\mathrm{wse}}_{2}$ heterobilayers},\ }\href {https://doi.org/10.1103/PhysRevLett.128.026402} {\bibfield  {journal} {\bibinfo  {journal} {Phys. Rev. Lett.}\ }\textbf {\bibinfo {volume} {128}},\ \bibinfo {pages} {026402} (\bibinfo {year} {2022})}\BibitemShut {NoStop}%
\bibitem [{\citenamefont {Lian}\ \emph {et~al.}(2023)\citenamefont {Lian}, \citenamefont {Meng}, \citenamefont {Ma}, \citenamefont {Maity}, \citenamefont {Yan}, \citenamefont {Wu}, \citenamefont {Huang}, \citenamefont {Chen}, \citenamefont {Chen}, \citenamefont {Chen}, \citenamefont {Blei}, \citenamefont {Taniguchi}, \citenamefont {Watanabe}, \citenamefont {Tongay}, \citenamefont {Lischner}, \citenamefont {Cui},\ and\ \citenamefont {Shi}}]{Lian23}%
  \BibitemOpen
  \bibfield  {author} {\bibinfo {author} {\bibfnamefont {Z.}~\bibnamefont {Lian}}, \bibinfo {author} {\bibfnamefont {Y.}~\bibnamefont {Meng}}, \bibinfo {author} {\bibfnamefont {L.}~\bibnamefont {Ma}}, \bibinfo {author} {\bibfnamefont {I.}~\bibnamefont {Maity}}, \bibinfo {author} {\bibfnamefont {L.}~\bibnamefont {Yan}}, \bibinfo {author} {\bibfnamefont {Q.}~\bibnamefont {Wu}}, \bibinfo {author} {\bibfnamefont {X.}~\bibnamefont {Huang}}, \bibinfo {author} {\bibfnamefont {D.}~\bibnamefont {Chen}}, \bibinfo {author} {\bibfnamefont {X.}~\bibnamefont {Chen}}, \bibinfo {author} {\bibfnamefont {X.}~\bibnamefont {Chen}}, \bibinfo {author} {\bibfnamefont {M.}~\bibnamefont {Blei}}, \bibinfo {author} {\bibfnamefont {T.}~\bibnamefont {Taniguchi}}, \bibinfo {author} {\bibfnamefont {K.}~\bibnamefont {Watanabe}}, \bibinfo {author} {\bibfnamefont {S.}~\bibnamefont {Tongay}}, \bibinfo {author} {\bibfnamefont {J.}~\bibnamefont {Lischner}}, \bibinfo {author} {\bibfnamefont {Y.-T.}\ \bibnamefont {Cui}},\ and\ \bibinfo {author} {\bibfnamefont {S.-F.}\ \bibnamefont {Shi}},\ }\bibfield  {title} {\bibinfo {title} {Valley-polarized excitonic {Mott} insulator in {WS2/WSe2} moir\'e superlattice},\ }\href {https://doi.org/10.1038/s41567-023-02266-2} {\bibfield  {journal} {\bibinfo  {journal} {Nature Physics}\ }\textbf {\bibinfo {volume} {20}},\ \bibinfo {pages} {34–39} (\bibinfo {year} {2023})}\BibitemShut {NoStop}%
\bibitem [{\citenamefont {Cai}\ \emph {et~al.}(2023)\citenamefont {Cai}, \citenamefont {Anderson}, \citenamefont {Wang}, \citenamefont {Zhang}, \citenamefont {Liu}, \citenamefont {Holtzmann}, \citenamefont {Zhang}, \citenamefont {Fan}, \citenamefont {Taniguchi}, \citenamefont {Watanabe}, \citenamefont {Ran}, \citenamefont {Cao}, \citenamefont {Fu}, \citenamefont {Xiao}, \citenamefont {Yao},\ and\ \citenamefont {Xu}}]{FQAH_MoTe2_Xu_2023a}%
  \BibitemOpen
  \bibfield  {author} {\bibinfo {author} {\bibfnamefont {J.}~\bibnamefont {Cai}}, \bibinfo {author} {\bibfnamefont {E.}~\bibnamefont {Anderson}}, \bibinfo {author} {\bibfnamefont {C.}~\bibnamefont {Wang}}, \bibinfo {author} {\bibfnamefont {X.}~\bibnamefont {Zhang}}, \bibinfo {author} {\bibfnamefont {X.}~\bibnamefont {Liu}}, \bibinfo {author} {\bibfnamefont {W.}~\bibnamefont {Holtzmann}}, \bibinfo {author} {\bibfnamefont {Y.}~\bibnamefont {Zhang}}, \bibinfo {author} {\bibfnamefont {F.}~\bibnamefont {Fan}}, \bibinfo {author} {\bibfnamefont {T.}~\bibnamefont {Taniguchi}}, \bibinfo {author} {\bibfnamefont {K.}~\bibnamefont {Watanabe}}, \bibinfo {author} {\bibfnamefont {Y.}~\bibnamefont {Ran}}, \bibinfo {author} {\bibfnamefont {T.}~\bibnamefont {Cao}}, \bibinfo {author} {\bibfnamefont {L.}~\bibnamefont {Fu}}, \bibinfo {author} {\bibfnamefont {D.}~\bibnamefont {Xiao}}, \bibinfo {author} {\bibfnamefont {W.}~\bibnamefont {Yao}},\ and\ \bibinfo {author} {\bibfnamefont {X.}~\bibnamefont {Xu}},\ }\bibfield  {title} {\bibinfo {title} {Signatures of fractional quantum anomalous {Hall} states in twisted {M}o{T}e2},\ }\href {https://doi.org/10.1038/s41586-023-06289-w} {\bibfield  {journal} {\bibinfo  {journal} {Nature}\ }\textbf {\bibinfo {volume} {622}},\ \bibinfo {pages} {63} (\bibinfo {year} {2023})}\BibitemShut {NoStop}%
\bibitem [{\citenamefont {Park}\ \emph {et~al.}(2023)\citenamefont {Park}, \citenamefont {Cai}, \citenamefont {Anderson}, \citenamefont {Zhang}, \citenamefont {Zhu}, \citenamefont {Liu}, \citenamefont {Wang}, \citenamefont {Holtzmann}, \citenamefont {Hu}, \citenamefont {Liu}, \citenamefont {Taniguchi}, \citenamefont {Watanabe}, \citenamefont {Chu}, \citenamefont {Cao}, \citenamefont {Fu}, \citenamefont {Yao}, \citenamefont {Chang}, \citenamefont {Cobden}, \citenamefont {Xiao},\ and\ \citenamefont {Xu}}]{FQAH_MoTe2_Xu_2023b}%
  \BibitemOpen
  \bibfield  {author} {\bibinfo {author} {\bibfnamefont {H.}~\bibnamefont {Park}}, \bibinfo {author} {\bibfnamefont {J.}~\bibnamefont {Cai}}, \bibinfo {author} {\bibfnamefont {E.}~\bibnamefont {Anderson}}, \bibinfo {author} {\bibfnamefont {Y.}~\bibnamefont {Zhang}}, \bibinfo {author} {\bibfnamefont {J.}~\bibnamefont {Zhu}}, \bibinfo {author} {\bibfnamefont {X.}~\bibnamefont {Liu}}, \bibinfo {author} {\bibfnamefont {C.}~\bibnamefont {Wang}}, \bibinfo {author} {\bibfnamefont {W.}~\bibnamefont {Holtzmann}}, \bibinfo {author} {\bibfnamefont {C.}~\bibnamefont {Hu}}, \bibinfo {author} {\bibfnamefont {Z.}~\bibnamefont {Liu}}, \bibinfo {author} {\bibfnamefont {T.}~\bibnamefont {Taniguchi}}, \bibinfo {author} {\bibfnamefont {K.}~\bibnamefont {Watanabe}}, \bibinfo {author} {\bibfnamefont {J.-H.}\ \bibnamefont {Chu}}, \bibinfo {author} {\bibfnamefont {T.}~\bibnamefont {Cao}}, \bibinfo {author} {\bibfnamefont {L.}~\bibnamefont {Fu}}, \bibinfo {author} {\bibfnamefont {W.}~\bibnamefont {Yao}}, \bibinfo {author} {\bibfnamefont {C.-Z.}\ \bibnamefont {Chang}}, \bibinfo {author} {\bibfnamefont {D.}~\bibnamefont {Cobden}}, \bibinfo {author} {\bibfnamefont {D.}~\bibnamefont {Xiao}},\ and\ \bibinfo {author} {\bibfnamefont {X.}~\bibnamefont {Xu}},\ }\bibfield  {title} {\bibinfo {title} {Observation of fractionally quantized anomalous {Hall} effect},\ }\href {https://doi.org/10.1038/s41586-023-06536-0} {\bibfield  {journal} {\bibinfo  {journal} {Nature}\ }\textbf {\bibinfo {volume} {622}},\ \bibinfo {pages} {74} (\bibinfo {year} {2023})}\BibitemShut {NoStop}%
\bibitem [{\citenamefont {Zeng}\ \emph {et~al.}(2023)\citenamefont {Zeng}, \citenamefont {Xia}, \citenamefont {Kang}, \citenamefont {Zhu}, \citenamefont {Kn{\"u}ppel}, \citenamefont {Vaswani}, \citenamefont {Watanabe}, \citenamefont {Taniguchi}, \citenamefont {Mak},\ and\ \citenamefont {Shan}}]{FQAH_MoTe2_Mak_Shan_2023}%
  \BibitemOpen
  \bibfield  {author} {\bibinfo {author} {\bibfnamefont {Y.}~\bibnamefont {Zeng}}, \bibinfo {author} {\bibfnamefont {Z.}~\bibnamefont {Xia}}, \bibinfo {author} {\bibfnamefont {K.}~\bibnamefont {Kang}}, \bibinfo {author} {\bibfnamefont {J.}~\bibnamefont {Zhu}}, \bibinfo {author} {\bibfnamefont {P.}~\bibnamefont {Kn{\"u}ppel}}, \bibinfo {author} {\bibfnamefont {C.}~\bibnamefont {Vaswani}}, \bibinfo {author} {\bibfnamefont {K.}~\bibnamefont {Watanabe}}, \bibinfo {author} {\bibfnamefont {T.}~\bibnamefont {Taniguchi}}, \bibinfo {author} {\bibfnamefont {K.~F.}\ \bibnamefont {Mak}},\ and\ \bibinfo {author} {\bibfnamefont {J.}~\bibnamefont {Shan}},\ }\bibfield  {title} {\bibinfo {title} {Thermodynamic evidence of fractional {Chern} insulator in moir{\'e} {M}o{T}e2},\ }\href {https://doi.org/10.1038/s41586-023-06452-3} {\bibfield  {journal} {\bibinfo  {journal} {Nature}\ }\textbf {\bibinfo {volume} {622}},\ \bibinfo {pages} {69} (\bibinfo {year} {2023})}\BibitemShut {NoStop}%
\bibitem [{\citenamefont {Xu}\ \emph {et~al.}(2023)\citenamefont {Xu}, \citenamefont {Sun}, \citenamefont {Jia}, \citenamefont {Liu}, \citenamefont {Xu}, \citenamefont {Li}, \citenamefont {Gu}, \citenamefont {Watanabe}, \citenamefont {Taniguchi}, \citenamefont {Tong}, \citenamefont {Jia}, \citenamefont {Shi}, \citenamefont {Jiang}, \citenamefont {Zhang}, \citenamefont {Liu},\ and\ \citenamefont {Li}}]{FQAHE_MoTe2_Li_2023}%
  \BibitemOpen
  \bibfield  {author} {\bibinfo {author} {\bibfnamefont {F.}~\bibnamefont {Xu}}, \bibinfo {author} {\bibfnamefont {Z.}~\bibnamefont {Sun}}, \bibinfo {author} {\bibfnamefont {T.}~\bibnamefont {Jia}}, \bibinfo {author} {\bibfnamefont {C.}~\bibnamefont {Liu}}, \bibinfo {author} {\bibfnamefont {C.}~\bibnamefont {Xu}}, \bibinfo {author} {\bibfnamefont {C.}~\bibnamefont {Li}}, \bibinfo {author} {\bibfnamefont {Y.}~\bibnamefont {Gu}}, \bibinfo {author} {\bibfnamefont {K.}~\bibnamefont {Watanabe}}, \bibinfo {author} {\bibfnamefont {T.}~\bibnamefont {Taniguchi}}, \bibinfo {author} {\bibfnamefont {B.}~\bibnamefont {Tong}}, \bibinfo {author} {\bibfnamefont {J.}~\bibnamefont {Jia}}, \bibinfo {author} {\bibfnamefont {Z.}~\bibnamefont {Shi}}, \bibinfo {author} {\bibfnamefont {S.}~\bibnamefont {Jiang}}, \bibinfo {author} {\bibfnamefont {Y.}~\bibnamefont {Zhang}}, \bibinfo {author} {\bibfnamefont {X.}~\bibnamefont {Liu}},\ and\ \bibinfo {author} {\bibfnamefont {T.}~\bibnamefont {Li}},\ }\bibfield  {title} {\bibinfo {title} {Observation of integer and fractional quantum anomalous {Hall} effects in twisted bilayer ${\mathrm{{m}o{t}e}}_{2}$},\ }\href {https://doi.org/10.1103/PhysRevX.13.031037} {\bibfield  {journal} {\bibinfo  {journal} {Phys. Rev. X}\ }\textbf {\bibinfo {volume} {13}},\ \bibinfo {pages} {031037} (\bibinfo {year} {2023})}\BibitemShut {NoStop}%
\bibitem [{Kun(2026)}]{Kundu_Balram_density_correlators_multiplet_repo}%
  \BibitemOpen
  \href {https://github.com/ritajitk/spin-polarized-fqh-energies} {\bibinfo {title} {Github repository: Exact relations between the density-density correlators of states in a spin multiplet}} (\bibinfo {year} {2026})\BibitemShut {NoStop}%
\bibitem [{Dia()}]{DiagHam}%
  \BibitemOpen
  \href@noop {} {}\bibinfo {note} {Diag{H}am, \url{https://www.nick-ux.org/diagham}}\BibitemShut {NoStop}%
\bibitem [{\citenamefont {Haldane}(1983)}]{Haldane83}%
  \BibitemOpen
  \bibfield  {author} {\bibinfo {author} {\bibfnamefont {F.~D.~M.}\ \bibnamefont {Haldane}},\ }\bibfield  {title} {\bibinfo {title} {Fractional quantization of the {Hall} effect: A hierarchy of incompressible quantum fluid states},\ }\href {https://doi.org/10.1103/PhysRevLett.51.605} {\bibfield  {journal} {\bibinfo  {journal} {Phys. Rev. Lett.}\ }\textbf {\bibinfo {volume} {51}},\ \bibinfo {pages} {605} (\bibinfo {year} {1983})}\BibitemShut {NoStop}%
\bibitem [{\citenamefont {Simon}\ and\ \citenamefont {Halperin}(1994)}]{Simon94a}%
  \BibitemOpen
  \bibfield  {author} {\bibinfo {author} {\bibfnamefont {S.~H.}\ \bibnamefont {Simon}}\ and\ \bibinfo {author} {\bibfnamefont {B.~I.}\ \bibnamefont {Halperin}},\ }\bibfield  {title} {\bibinfo {title} {Response function of the fractional quantized {Hall} state on a sphere. i. fermion {Chern}-{Simons} theory},\ }\href {https://doi.org/10.1103/PhysRevB.50.1807} {\bibfield  {journal} {\bibinfo  {journal} {Phys. Rev. B}\ }\textbf {\bibinfo {volume} {50}},\ \bibinfo {pages} {1807} (\bibinfo {year} {1994})}\BibitemShut {NoStop}%
\bibitem [{\citenamefont {He}\ \emph {et~al.}(1994)\citenamefont {He}, \citenamefont {Simon},\ and\ \citenamefont {Halperin}}]{He94}%
  \BibitemOpen
  \bibfield  {author} {\bibinfo {author} {\bibfnamefont {S.}~\bibnamefont {He}}, \bibinfo {author} {\bibfnamefont {S.~H.}\ \bibnamefont {Simon}},\ and\ \bibinfo {author} {\bibfnamefont {B.~I.}\ \bibnamefont {Halperin}},\ }\bibfield  {title} {\bibinfo {title} {Response function of the fractional quantized {Hall} state on a sphere. ii. exact diagonalization},\ }\href {https://doi.org/10.1103/PhysRevB.50.1823} {\bibfield  {journal} {\bibinfo  {journal} {Phys. Rev. B}\ }\textbf {\bibinfo {volume} {50}},\ \bibinfo {pages} {1823} (\bibinfo {year} {1994})}\BibitemShut {NoStop}%
\bibitem [{\citenamefont {Girvin}\ \emph {et~al.}(1985)\citenamefont {Girvin}, \citenamefont {MacDonald},\ and\ \citenamefont {Platzman}}]{Girvin85}%
  \BibitemOpen
  \bibfield  {author} {\bibinfo {author} {\bibfnamefont {S.~M.}\ \bibnamefont {Girvin}}, \bibinfo {author} {\bibfnamefont {A.~H.}\ \bibnamefont {MacDonald}},\ and\ \bibinfo {author} {\bibfnamefont {P.~M.}\ \bibnamefont {Platzman}},\ }\bibfield  {title} {\bibinfo {title} {Collective-excitation gap in the fractional quantum {Hall} effect},\ }\href {https://doi.org/10.1103/PhysRevLett.54.581} {\bibfield  {journal} {\bibinfo  {journal} {Phys. Rev. Lett.}\ }\textbf {\bibinfo {volume} {54}},\ \bibinfo {pages} {581} (\bibinfo {year} {1985})}\BibitemShut {NoStop}%
\end{thebibliography}%

\end{document}